\documentclass[twocolumn,
showpacs,amsfonts,amssymb,amsmath,floats,superscriptaddress,aps,longbibliography]{revtex4-1}

\usepackage{graphicx}
\usepackage{upgreek}
\usepackage{color}
\usepackage[section]{placeins}
\usepackage[colorlinks=true]{hyperref}
\usepackage[english]{babel}
\usepackage{bm}
\usepackage{braket}

\DeclareMathOperator{\arccot}{arccot}
\DeclareMathOperator{\artanh}{artanh}
\DeclareMathOperator{\tr}{Tr}

\def\mr#1{\mathrm{#1}}

\def\up{\uparrow}
\def\down{\downarrow}

\def\a0{A_0}
\def\wh{\omega_h}

\def\be{\beta_e}
\def\bn{\beta_n}

\usepackage[normalem]{ulem}

\begin{document}

\title{Nuclear spin polaron-formation: anisotropy effects and quantum phase transition}

\author{Iris Kleinjohann}
\affiliation{Condensed Matter Theory, Department of Physics, TU Dortmund University,
44221 Dortmund, Germany}

\author{Andreas Fischer}
\affiliation{Condensed Matter Theory, Department of Physics, TU Dortmund University,
44221 Dortmund, Germany}

\author{Mikhail M.\ Glazov}
\affiliation{Ioffe Institute, 194021 St.Petersburg, Russia}

\author{Frithjof B.\ Anders}
\affiliation{Condensed Matter Theory, Department of Physics, TU Dortmund University,
44221 Dortmund, Germany}

\date{\today}

\begin{abstract}

We study theoretically the formation of the nuclear-spin polaron state in semiconductor nanosystems within the Lindblad equation approach.
To this end, we derive a general Lindblad equation for the density operator that complies with the symmetry of the system Hamiltonian and address the nuclear-spin polaron formation for localized charge carriers subject to an arbitrarily anisotropic hyperfine interaction when optically cooling the nuclei.
The steady-state solution of the density matrix for an anisotropic central spin model is presented as a function of the electron and nuclear spin bath temperature.
Results for the electron-nuclear spin correlator as well as data for the nuclear spin distribution function serve as a measure of spin-entanglement.
The features in both of them clearly indicate the formation of the nuclear polaron state at low temperatures where the crossover regime coincides with an enhancement of quantum fluctuations and agrees with the mean-field prediction of the critical temperature line.
We can identify two distinct polaron states dependent upon the hyperfine anisotropy which are separated by a quantum phase transition at the isotropic point.
These states are reflected in the temporal spin auto-correlation functions accessible in experiment via spin-noise measurements.

\end{abstract}

\maketitle

\section{Introduction}

The investigation of the electron spin dynamics in semiconductor quantum dots (QDs) has caused a very large interest in the last two decades \cite{HansonSpinQdotsRMP2007,dyakonov_book,glazov_book,smirnov:SNS:rev} due to the magnificent fundamental physics and the possible applications in quantum technologies.
The entanglement in interacting spin systems is of high relevance nowadays~\cite{PhysRevLett.107.206806,Gangloff2019,Gangloff:2021te,PhysRevLett.126.216804}.
In particular, the entanglement between the electron and nuclear spins is mediated by the hyperfine interaction between the locally bound charge carrier spin and the surrounding nuclear spins that limits the electron spin coherence time \cite{merkulov_prb2002} in QDs with disordered nuclear spins.
While the fluctuating Overhauser field acting on the electron from the disordered nuclear spins is only of the order of $10$~mT, 
polarized nuclei can generate an effective magnetic field of several Tesla in GaAs-type semiconductors~\cite{merkulov_pss1998,dyakonov_book,glazov_book}.

The electron spin affects the nuclei via the Knight field induced by the hyperfine interaction and can be efficiently oriented optically~\cite{opt_or_book,dyakonov_book,glazov_book}.
As a result, optical excitation is responsible for the dynamic nuclear polarization in InAs/GaAs QDs \cite{eble_prb2006} as well as mode locking \cite{greilich_science2006} and nuclei-induced frequency focusing effects \cite{GreilichBayer2007,Evers2018} enabling efficient control of the nuclear spin degrees of freedom by non-magnetic means.

When lowering the temperature, the correlated ground state of the system becomes dominant: 
electron and nuclear spins corroborate and form a correlated or entangled nuclear-spin polaron state that minimizes the hyperfine energy.
Such a state has been predicted by Merkulov \cite{merkulov_pss1998} in a framework of the mean-field quasi equilibrium model, assigning the electron and nuclear spins different effective temperatures.
The two temperatures, $T_e$ and $T_n$, were used in mean-field theory \cite{merkulov_pss1998} to predict a critical temperature line on which the transition from an uncorrelated system to a nuclear-polaronic state occurs.
The key idea is based on the observation that the electron remains coupled to the lattice, whereas the very long lifetime of the nuclear spin polarization up to several hours \cite{kkm_nucl_book,vladimirova_prb2017} indicates a strong decoupling of the nuclear spins from the environment. 
While the electronic degrees of freedom maintain their base temperature $T_e$ (typically, on the order of several Kelvin), the spin temperature $T_n$ of optically cooled nuclei can be much lower than $T_e$~\cite{opt_or_book,dyakonov_book,glazov_book,Chekhovich2017,vladimirova_prb2018,Kotur:2021wp}.
In particular, recently Ref.~\cite{Kotur:2021wp} reported a nuclear spin temperature as low as $0.54$~$\mu$K.

Progress in the cooling of the nuclear spin systems motivates theoretical studies of the entangled electron-nuclear spin states.
The analysis of the nuclear-spin polaron formation beyond the mean-field approach was presented in Ref.~\cite{scalbert_prb2017}.
In Ref.~\cite{PhysRevB.103.205207}, in addition to the nuclear-spin polaron, a novel state termed a dynamically induced nuclear ferromagnet was predicted.
In a recent paper \cite{fischer_prb2020}, we explored the nuclear polaron formation beyond the mean-field theory by employing a master equation for the distribution function of the interacting electron-nuclear spin system.
The analysis in Ref.~\cite{fischer_prb2020} was restricted to the Ising limit
of the hyperfine interaction, where the eigenstates of the system can be conveniently expressed as products of the electron and nuclear spin states and the spin-flip transition rates between those states mediated by the coupling with external reservoirs can be explicitly written.
The solution of the corresponding master equation has made it possible to obtain not only the transition temperature to the nuclear-spin polaron state, but also the distribution functions of the spins, the fluctuations of electron and nuclear spins and address the dynamics of the polaron formation.
In this paper, we substantially extend the theory to investigate the polaronic state for an arbitrary anisotropic hyperfine interaction, needed to access the physical relevant regimes in semiconductor QDs where an isotropic hyperfine interaction is realized for electrons, Ising-like interaction for the heavy-holes, and anisotropic interaction for the light-holes and heavy-light hole mixtures~\cite{glazov_book}.

We derive a generalized Lindblad approach to two spin reservoirs that impose the two temperatures, $T_e$ and $T_n$, as boundary conditions.
Our approach is suitable for all temperature regimes, and the Lindblad rates are fixed in such a way that the steady state solution of the Lindblad equation is given by the Boltzmann form of the density matrix in thermal equilibrium.
In order to address the nuclear polaron formation in a system with a very large number of nuclear spins upto $N=1000$ in a semi-analytical fashion, we resort to the box model approximation \cite{RS1983,1742-5468-2007-06-P06018,kozlov_jetp2007} of the central spin model (CSM).
We investigate the nuclear polaron formation as a function of the anisotropy parameter $\lambda$ \cite{fischer_prb2008,hackmann_prb2014} where the limit $\lambda=0$ corresponds to the Ising limit \cite{fischer_prb2020} relevant for a purely heavy-hole bound QD state, $\lambda=1$ to the isotropic case of a negatively charged QD, and $\lambda>1$ to the regime of a mixture of heavy and light holes.
This allows to study all relevant regimes of positively and negatively charged InGaAs QDs.

We show that the polaron state is not destroyed by the quantum fluctuations present when reducing the nuclear bath temperature. 
The crossover regime is very narrow and follows the mean-field approach to the anisotropic CSM \cite{Gaudin1976,merkulov_pss1998,coish_prb2004,glazov_book}. 
In the absence of a symmetry breaking field, however, the nuclear polaronic state still contains the full degeneracy of the ground state in contrary to the mean-field theory. 

The paper is organized as follows.
Section~\ref{sec:model} is devoted to the presentation of our Lindblad approach where the included Lindblad operators mediate spin excitations caused by the coupling to the thermal reservoirs.
A general Hamiltonian for the hyperfine interaction is introduced in Sec.\ \ref{sec:hf} and the related Lindblad equation is presented in Sec.\ \ref{sec:lindblad}.
The rate equations for the density matrix in the energy eigenbasis are deduced in Sec.\ \ref{sec:elementwise}.
We adopt the general approach to the anisotropic CSM in Sec.\ \ref{sec:smodel}.
After the model is defined in Sec.\ \ref{sec:bm} and the box model eigenstates \cite{kozlov_jetp2007} are presented, the question of the determination of the Lindblad decay rates is addressed in Sec.\ \ref{sec:bmlindblad}. 
Section \ref{sec:polaronstate} is devoted to the emerging nuclear-spin polaron state.
We begin with the presentation of the electron-nuclear spin correlators as a function of temperature for different anisotropy parameters $\lambda$ in Sec.\ \ref{sec:nuclear-polaron-correlators} and compare our stationary Lindblad solution with a simplifying mean-field approach in Sec.\ \ref{sec:mf}.
The critical temperature of the polaron formation and the quantum fluctuations close to the very narrow crossover region are discussed in Sec.\ \ref{sec:temperature-crit}.
We address the nuclear spin distribution in Sec.\ \ref{sec:nucldistfunc} by tracing out the electronic spin configuration.
Our results are linked to a quantum phase transition that occurs at the isotropy point $\lambda=1$.
We discuss the change of the ground state at the critical point in Sec.\ \ref{sec:quantum-phase-transition}.
In Sec.~\ref{sec:fluct}, we present calculations for the spin auto-correlation function of the open quantum system.
Section~\ref{sec:szsz} is devoted to the real time dynamics of the electron spin and Sec.\ \ref{sec:jzjz} extends the discussion to the fluctuations of the nuclear spins.
We finish the paper with a short conclusion.

\section{Model}
\label{sec:model}

In this paper we investigate the formation of a polaronic state and its properties in a system with one localized electronic charge.
We explicitly treat the interaction between the nuclear spins and the localized charge carrier spin via the central spin model (CSM) and include energy and spin exchange with reservoirs within a set of Markovian transition rates.
We start with a presentation of the basic formalism.

\subsection{Hyperfine interaction}
\label{sec:hf}

The hyperfine interaction between the localized charge carrier spin $\mathbf{S}$ and the surrounding nuclear spins $\mathbf{I}_k$ is described by the Hamiltonian \cite{dyakonov_book,glazov_book}
\begin{equation}
H = \sum_{k=1}^N \sum_{\alpha,\beta} A_k^{\alpha,\beta} S^\alpha I_k^\beta .
\label{eq:H}
\end{equation}
Here we label the individual nuclear spins with an index $k \in \left\{ 1, \ldots , N \right\}$ and include all nuclear spins within the charge carrier localization volume.
The matrix $A_k^{\alpha,\beta}$ defines the generally anisotropic hyperfine coupling strength of an individual nuclear spin;
its matrix elements incorporate the electron wave function at the position of the respective nucleus, where $\alpha$ and $\beta \in \left\{ x,y,z \right\}$ refer to the Cartesian axes.

The Hamiltonian, Eq.~\eqref{eq:H}, accounts for a system with an anisotropic hyperfine coupling as well as the isotropic case, where, naturally, $A_k^{\alpha,\beta} \propto \delta_{\alpha,\beta}$ and $\delta_{x,y}$ is the Kronecker $\delta$-symbol \cite{glazov_book}.
Hamiltonian~\eqref{eq:H} is applicable to the description to a variety of semiconductor nanostructures such as singly charged QDs \cite{merkulov_prb2002,abragam_book} or donor-bound electrons \cite{feher_pr1959,pla_nature2012}.
Generally, the charge carrier spin $\mathbf{S}$ can portray an electron spin or a light/heavy hole spin involving a proper adjustment of the spin length and the hyperfine coupling constants $A_k^{\alpha,\beta}$ \cite{abragam_book,testelin_prb2009,hackmann_prl2015}.

\subsection{Lindblad formalism for thermal reservoirs}
\label{sec:lindblad}

To account for the effect of the optical cooling of the nuclear spin bath, we introduce a two-temperature concept \cite{opt_or_book,dyakonov_book,vladimirova_prb2018} with distinct effective inverse temperatures for the electron spin, $\be = 1/k_B T_e$, and the nuclear spins, $\bn = 1/k_B T_n$ \cite{glazov_book,merkulov_pss1998,scalbert_prb2017,fischer_prb2020}.
Under optical cooling of the nuclear spin bath, the electron spin mostly retains the lattice temperature while the nuclear spins are cooled below, $\bn > \be$.

Consequently, we treat the system as an open quantum system whose dynamics is driven by a unitary time evolution provided by the Hamiltonian $H$, Eq.~\eqref{eq:H}, and some Markovian transition rates between the eigenstates that account for the reservoir inducted energy and spin exchange.
Formally, this can be done by introducing fluctuating effective magnetic fields induced by reservoirs and acting on the electron and nuclear spins~\cite{fischer_prb2020}.
Corresponding coherent and incoherent dynamics of the system is most conveniently described by the density matrix.
Its evolution is governed by the Lindblad master equation \cite{carmichael_book}.

To that end, it is useful to introduce the complete eigenbasis of $H$ in Eq.~\eqref{eq:H}, as $H\ket{\psi_n} = \epsilon_n \ket{\psi_n}$, with eigenenergies $\epsilon_n$ and eigenvectors $\ket{\psi_n}$; the subscript $n$ enumerates all basic states of the systems.
The eigenbasis is used to define the complete operator basis $X_{mn} = \ket{\psi_m} \bra{\psi_n}$ of the Hilbert space.
Taking into account likely degeneracies of the eigenstates, the most general Lindblad operators $L_{m,n}^{k,\alpha}$ in the form
\begin{eqnarray}
L_{m,n}^{k,\alpha} &=& \sqrt{\Gamma_{m,n}^{k,\alpha}} \sum_{a,b} \delta_{\epsilon_a,\epsilon_m} \delta_{\epsilon_b,\epsilon_n} \braket{\psi_a|s_k^\alpha|\psi_b} X_{ab},
\label{eq:Lb} 
\end{eqnarray}
describe transitions between the eigenstates $\ket{\psi_n}$ and $\ket{\psi_m}$ that are mediated by the reservoirs with the rate $\Gamma_{m,n}^{k,\alpha}$ (presented below) via the spin-operator $s_k^\alpha$.

>From now on,  the index $k$ refers to either the electron spin ($k=0$), $s_0^\alpha=S^\alpha$, or one of the nuclear spins ($k \in \left\{ 1, \ldots , N \right\}$), $s_k^\alpha=I_k^\alpha$ for convenience.
The sum over all states $a,b$ accounts for all combinations of initial and final states sharing the same transition energy difference 
\begin{equation}
\label{trans:energ}
\Delta_{mn}=\epsilon_m - \epsilon_n,
\end{equation}
due to the degeneracy of states. 
These Lindblad operators and their Hermitian conjugates, $(L_{m,n}^{k,\alpha})^\dag$, enter the Lindblad master equation,
\begin{align}
\dot{\rho} = \mathcal{L} \rho &=
-i \left[ H, \rho \right] - \sum_{k=0}^N \sum_\alpha \sum_{m,n} \left\{ (L_{m,n}^{k,\alpha})^\dag L_{m,n}^{k,\alpha} \rho \right. \notag \\
&\qquad \left. + \rho (L_{m,n}^{k,\alpha})^\dag L_{m,n}^{k,\alpha} - 2 L_{m,n}^{k,\alpha} \rho (L_{m,n}^{k,\alpha})^\dag \right\} ,
\label{eq:La} 
\end{align}
governing the temporal evolution of the system's density operator $\rho$.

Generally, the transition rates must be constructed in such a way that the steady-state solution of the density operator in thermal equilibrium aquires the Boltzmann form which commutes with $H$.
Accordingly, the rate of a respective transition is given by 
\begin{eqnarray}
\Gamma_{m,n}^{k,\alpha} &=& \frac{W_k^\alpha h_k^\alpha(\Delta_{mn})}{g(\epsilon_m) g(\epsilon_n)},
\label{eq:gamma-rate}
\end{eqnarray}
where $g(\epsilon_m)$ denotes the degeneracy of the eigenenergy $\epsilon_m$ and $W_k^\alpha$ some phenomenological rate that typically is assumed to be several orders of magnitude larger for the electron spin than for the nuclear spins due to the electron's stronger coupling to the environment.
The usefulness of separation between the rate $W_k^\alpha$ and the degeneracy factor $g(\epsilon_m)$ becomes clear below in Sec.\ \ref{sec:elementwise}.

The dimensionless function $h_k^\alpha(\Delta_{mn})$ takes into account an enhancement or suppression of transitions depending on the energy difference between the initial and final states, Eq.~\eqref{trans:energ}.
Demanding the relaxation of $\rho$ to the Boltzmann form in thermodynamic equilibrium requires the ratio
$h_k^\alpha(\Delta_{mn})/h_k^\alpha(-\Delta_{mn}) = \exp (-\Delta_{mn} \beta_k)$, where $\beta_k=\beta$.
In this paper, we allow for two different effective inverse spin reservoir temperatures $\beta_k=\beta_e$ for $k=0$ and $\beta_k=\beta_n$ otherwise as it takes place in the experiments on the optical cooling of lattice nuclei~\cite{opt_or_book,glazov_book,vladimirova_prb2018}.

The above formulation, Eqs.~\eqref{eq:Lb} and \eqref{eq:La}, of the two-reservoir concept for the electron-nuclear spin system constitutes an extension of the rate-equation formalism introduced in Ref.~\cite{fischer_prb2020}.
The Lindblad equation incorporates off-diagonal elements of the density operator $\rho$ and thereby allows for the description of the hyperfine interaction beyond the Ising limit.
For the Ising limit of the hyperfine coupling constants $A_k^{\alpha,\beta}$, it reproduces the results in Ref.~\cite{fischer_prb2020} as a special case.

However, the inclusion of the off-diagonal elements of $\rho$ facilitates the treatment of observables where the corresponding quantum mechanical operator does not commute with the Hamiltonian.
Therefore, this approach goes well beyond the previously considered Ising limit and pushes the theory into  experimentally relevant realms.

\subsection{Dynamics of  the density matrix}
\label{sec:elementwise}

In the definition of the Lindblad operator, Eq.~\eqref{eq:Lb}, the pair of sums over the energy eigenstates $a$ and $b$ in combination with the Kronecker $\delta$-symbols allows for contributions only from the eigenstates $\ket{\psi_a}$ ($\ket{\psi_b}$ respectively) that belong to the same energetically degenerate subspace as the state $m$ ($n$), i.\ e.\ the states for which $\epsilon_a = \epsilon_m$ ($\epsilon_b = \epsilon_m$).
In case of non degenerate eigenenergies, these sums reduce to a single contribution.
For degenerate eigenenergies however this construction ensures a free choice of the orthonormal eigenbasis within the energetically degenerate subspaces without altering the dynamics.
To avoid a double counting of the transitions, we include the degree of degeneracies $g(\epsilon_m)$, $g(\epsilon_n)$ as a prefactor in Eq.~\eqref{eq:gamma-rate}.
The details of the analysis are presented in Appendix~\ref{app:degen}.

To obtain the coupled differential equations for the density matrix, we convert Eq.~\eqref{eq:La}, see also Eqs. \eqref{eq:Lc}, \eqref{eq:Ld}, to a matrix representation using the energy eigenstates of $H$ and arrive at
\begin{multline}
\dot{\rho}_{mn} =-i \Delta_{mn} \rho_{mn} \\
- \sum_{k,\alpha} W_k^\alpha\sum_{a,b} 
\Bigg\{ \delta_{\epsilon_m,\epsilon_b} h_k^\alpha(\Delta_{am})
\left(s_k^\alpha \right)^*_{a,m}
\left( s_k^\alpha \right)_{a,b} \rho_{bn} \Bigg. \\
+ \delta_{\epsilon_n,\epsilon_{b}} h_k^\alpha(\Delta_{a n})
\left(s_k^\alpha \right)^*_{a,b}
\left( s_k^\alpha\right)_{a,n} \rho_{m b}\\
 \Bigg. -2 \delta_{\epsilon_m,\epsilon_n}
\delta_{\epsilon_a,\epsilon_{b}}
h_k^\alpha(\Delta_{m a})
\left(s_k^\alpha \right)_{m,a} \left(s_k^\alpha\right)^*_{n,b}
\rho_{ab} \Bigg\} .
\label{eq:Le}
\end{multline}
This equation can be conveniently used for numerical calculations.

\section{Models of hyperfine coupling and transition rates}
\label{sec:smodel}

Here, the general description for an arbitrary hyperfine coupling Hamiltonian, Eq.~\eqref{eq:H}, is customized to a more specific system where the hyperfine interaction anisotropy is uniaxial and described by a single parameter $\lambda$.
The corresponding master equation taking into account the coupling to thermal reservoirs is derived from general Eqs.~\eqref{eq:La} and \eqref{eq:Le}.

\subsection{Anistropic central spin model}
\label{sec:bm}

In systems such as singly charged self-assembled GaAs-type QDs grown on the $(xy) \parallel (001)$ crystallographic plane, the matrix $A_k^{\alpha,\beta}$ describing the hyperfine interaction, Eq.~\eqref{eq:H}, is diagonal and the coupling is, as a rule, isotropic in the $(xy)$ plane~\cite{glazov_book}.
The resulting Hamiltonian,
\begin{equation}
\label{eq:H:uni}
H = \sum_k A_k [ \lambda ( S^x I^x_k + S^y I^y_k ) + S^z I^z_k ] ,
\end{equation}
includes a uniaxial anisotropy parameter $\lambda$ with respect to the $z\parallel [001]$ direction. 
The Hamiltonian Eq.~\eqref{eq:H:uni} allows for the description of a variety of semiconductor nanostructures, although the physical origin of the coupling $A_k$ might differ.
The analysis of the situation with biaxial anisotropy or non-collinear hyperfine interaction~\cite{glazov_book,PhysRevB.94.121302,avdeev_nanoad2019} can be performed in the same way and goes beyond the scope of the present paper.

We recall that for the conduction band electron in an $s$-type orbital at an atomic site, the main contribution to the hyperfine coupling stems from the Fermi contact interaction \cite{fermi_zp1930}.
In contrast, for a hole spin coupling to the surrounding nuclear spins, the Fermi contact coupling is strongly suppressed due to the $p$-type wave function, and the dipole-dipole interaction is predominant \cite{testelin_prb2009}.
The coupling strength of the respective scenario is adjusted by the constants $A_k$ and the anisotropy is respected by the parameter $\lambda$ \cite{testelin_prb2009,fischer_prb2008,hackmann_prl2015}.
For $\lambda =1$, the isotropic limit relevant for an electron spin is restored whereas $\lambda=-2$ is a typical parameter for the spin of a light hole.
The Ising limit, $\lambda =0$, captures the heavy hole in a self-assembled InAs/GaAs QD with the sample's growth direction matching the $z$ axis.
In QDs the hole state often is a mixture of the heavy and light hole contribution depending on the geometry of the dot.
In such a case, the coupling can be described by the Hamiltonian~\eqref{eq:H:uni} with the parameter $\lambda$ varying, typically, between $-2$ and $0$.

To enable analytic access to the eigenenergies and eigenstates of the hyperfine Hamiltonian with a relatively large number of nuclear spins, $N\approx 1000$, we set the hyperfine coupling constant $A_k = A_0$ for all nuclear spins which is referred to as the box model approximation.
In this case the Hamiltonian can be written in terms of the total nuclear spin $\mathbf{J} = \sum_k \mathbf{I}_k$, 
\begin{equation}
\begin{split}
H &= A_0 \left[ \lambda \left( S^x J^x + S^y J^y \right) + S^z J^z \right] \\
&= A_0 \left[ \frac{\lambda}{2} \left( S^+ J^- + S^-J^+ \right) + S^z J^z \right]
\end{split}
\label{eq:bm}
\end{equation}
with the ladder operators of the electron spin $S^\pm = S^x \pm i S^y$ and the total nuclear spin $J^\pm = J^x \pm i J^y$.
As a characteristic frequency scale of the system we introduce $\omega_h = ( \sum_k A_k^2 )^{1/2} \equiv \sqrt{N} A_0$ based on the dephasing rate of the electron spin in the nuclear spin bath for $\lambda=1$.
We employ $\omega_h$ as a reference scale, e.\ g., for indicating energies and temperatures, in the following.

Since only the total nuclear spin $J$ and the quantum number $J^z$ enter the determination of the eigenstates, we distinguish between the different degenerate multiples \cite{kozlov_jetp2007} arising from the addition theorem for spin with the same $J$ by the index $\gamma$.
The eigenenergies $\epsilon_{J,J^z}^\sigma$ and eigenstates $\ket{\psi^{\sigma,\gamma}_{J,J^z}}$ for a system, in which the central spin $\mathbf{S}$ and the individual nuclear spins $\mathbf{I}_k$ have a length $1/2$ respectively, have been calculated by Kozlov \cite{kozlov_jetp2007} and read
\begin{subequations}
\label{eq:bmmmm}
\begin{align} 
\epsilon^+_{J,-J}  &= \frac{A_0 J}{2},
&\epsilon^+_{J,J+1} &= \frac{A_0 J}{2}, \label{eq:bm1e}\\
\ket{\psi^{+,\gamma}_{J,-J}} &= \ket{\downarrow} \ket{J,-J,\gamma},
&\ket{\psi^{+,\gamma}_{J,J+1}} &= \ket{\uparrow} \ket{J,J,\gamma}, \label{eq:bm1s}
\end{align}
\end{subequations}
with $J \in \left\{ 0,\ldots,N/2 \right\}$ and
\begin{subequations}
\label{eq:spectrum:boxK}
\begin{align}
\epsilon^\pm_{J,J^z} &= - \frac{A_0}{4} \pm \frac{A_0}{2} \Bigg\{ \left(J^z-\frac12 \right)^2 \Bigg. \notag \\
& \qquad \Bigg. + \lambda^2 \left[ J(J+1) -J^z(J^z-1) \right] \Bigg\}^{1/2}
\label{eq:bm2e} \\
\ket{\psi^{\sigma,\gamma}_{J,J^z}} &= c^\sigma_{J,J^z} \ket{\downarrow} \ket{J,J^z,\gamma} + d^\sigma_{J,J^z} \ket{\uparrow} \ket{J,J^z-1,\gamma} \label{eq:bm2s}
\end{align}
\end{subequations}
where $J \in \left\{ 0,\ldots,N/2 \right\}$, $J^z \in \left\{ -J+1,\ldots,J \right\}$ and $\sigma \in \left\{ +,- \right\}$. 
The eigenstates are given in terms of the electron spin and the total nuclear spin $z$ product basis with $\ket{\uparrow / \downarrow}$ referring to the electron spin state and $\ket{J,J^z,\gamma}$ determining the nuclear spin state with the quantum numbers for total nuclear spin length $J$ and the $z$ quantum number $J^z$.

The coefficients $c^\sigma_{J,J^z}$ and $d^\sigma_{J,J^z}$ of the eigenstates, Eq.~\eqref{eq:bm2s}, are obtained from analytical diagonalization of the $2\times 2$ dimensional subblocks of the Hamilton matrix spanned by the states $\ket{\down} \ket{J,J^z,\gamma}$ and $\ket{\up} \ket{J,J^z-1,\gamma}$,
\begin{align}
H_{J,J^z}^{2\times2} =
\begin{pmatrix}
-A_0 J^z/2 & T_{J,J^z} \\
T_{J,J^z} & A_0(J^z-1)/2
\end{pmatrix},
\end{align}
with $T_{J,J^z} = \lambda A_0 \sqrt{J(J+1)-J^z(J^z-1)}$.
Note that the label $J^z=J+1$ in the Eqs.~\eqref{eq:bm1e} and \eqref{eq:bm1s} does not correspond to the actual quantum number of the state, but is chosen in compliance with the labeling in Eqs.~\eqref{eq:bm2e} and \eqref{eq:bm2s}, and allows for a general notation of eigenenergies $\epsilon^\sigma_{J,J^z}$ and eigenstates $\ket{\psi^{\sigma,\gamma}_{J,J^z}}$ where $J^z \in \left\{ -J,\ldots,J+1 \right\}$.

As mentioned above, the quantity $\gamma$ accounts for the degeneracy in the system since the Hamilton matrix is block diagonal and can be split into subblocks with fixed quantum number $J$ whereby for each value of $J$ a number $g_N(J)$ of identical blocks exist.
Assuming an even number $N$ of nuclear spins, this degree of degeneracy is given by
\begin{equation}
g_N(J) = \frac{2J+1}{N/2+J+1} {N\choose{N/2+J}},
\label{eq:Jdeg}
\end{equation}
where ${a\choose b} = a!/[b!(a-b)!]$ is the binomial coefficient.

\subsection{Reduced rate equations}
\label{sec:bmlindblad}

With the aid of the eigenstate decomposition, Eqs.~\eqref{eq:bmmmm} and \eqref{eq:spectrum:boxK}, we specify the final master equation in the box model limit:
Each sum over the eigenstates in the original master equation, Eq.~\eqref{eq:Le}, is split into sums over the box model quantum numbers, $J$, $J^z$, $\sigma$, and $\gamma$. Furthermore, we can assume the density operator to be diagonal in the quantum numbers $J$ and $\gamma$ as the Hamiltonian and thereby reduce the number of sums.
Next, we replace the operator $s_k^\alpha$ in Eq.~\eqref{eq:Le} by a ladder operator, $s_k^\tau$
\begin{equation}
s_k^\tau = \begin{cases} s_k^+ / \sqrt{2}, & \tau = -1, \\
s_k^z, & \tau = 0, \\
s_k^- / \sqrt{2}, & \tau = +1, \end{cases}
\label{eq:sktau}
\end{equation}
with the factor $1/\sqrt{2}$ stemming from normalization.

Taking into account that a spin-flip element $\braket{\psi^{\sigma',\gamma'}_{J',{J^z}'}|s_k^\tau|\psi^{\sigma,\gamma}_{J,J^z}}$ only yields a contribution when ${J^z}' = J^z+\tau$, independent on the fact which spin $k$ is flipped, one obtains the master equation for the density matrix elements
\begin{multline}
\partial_t \braket{\psi^{\sigma_m,\gamma}_{J,J^z_m}|\rho|\psi^{\sigma_n,\gamma}_{J,J^z_n}} = 
-i \Delta^{\sigma_m,J,J^z_m}_{\sigma_n,J,J^z_n}
\braket{\psi^{\sigma_m,\gamma}_{J,J^z_m}|\rho|\psi^{\sigma_n,\gamma}_{J,J^z_n}} \\
- \sum_{k,\tau} W_k^\tau
\sum_{J',\gamma'} \sum_{\sigma,\sigma'} \Bigg\{ \delta_{\epsilon^{\sigma_m}_{J,J^z_m},\epsilon^{\sigma'}_{J,J^z_m}}
h_k^\tau(\Delta^{\sigma,J',J^z_m+\tau}_{\sigma_m,J,J^z_m}) \Bigg. \\
\braket{\psi^{\sigma_m,\gamma}_{J,J^z_m}|(s_k^\tau)^\dag|\psi^{\sigma,\gamma'}_{J',J^z_m+\tau}}
\braket{\psi^{\sigma,\gamma'}_{J',J^z_m+\tau}|s_k^\tau|\psi^{\sigma',\gamma}_{J,J^z_m}} \\
\braket{\psi^{\sigma',\gamma}_{J,J^z_m}|\rho|\psi^{\sigma_n,\gamma}_{J,J^z_n}}
+ \delta_{\epsilon^{\sigma_n}_{J,J^z_n},\epsilon^{\sigma'}_{J,J^z_n}}
h_k^\tau(\Delta^{\sigma,J',J^z_n+\tau}_{\sigma_n,J,J^z_n}) \\
\braket{\psi^{\sigma',\gamma}_{J,J^z_n}|(s_k^\tau)^\dag|\psi^{\sigma,\gamma'}_{J',J^z_n+\tau}}
\braket{\psi^{\sigma,\gamma'}_{J',J^z_n+\tau}|s_k^\tau|\psi^{\sigma_n,\gamma}_{J,J^z_n}} \\
\braket{\psi^{\sigma_m,\gamma}_{J,J^z_m}|\rho|\psi^{\sigma',\gamma}_{J,J^z_n}}
-2 \delta_{\epsilon^{\sigma_m}_{J,J^z_m},\epsilon^{\sigma_n}_{J,J^z_n}} \delta_{\epsilon^\sigma_{J',J^z_m-\tau},\epsilon^{\sigma'}_{J',J^z_n-\tau}} \\
h_k^\tau(\Delta^{\sigma_m,J,J^z_m}_{\sigma,J',J^z_m-\tau})
\braket{\psi^{\sigma_m,\gamma}_{J,J^z_m}|s_k^\tau|\psi^{\sigma,\gamma'}_{J',J^z_m-\tau}} \\
\Bigg. \braket{\psi^{\sigma',\gamma'}_{J',J^z_n-\tau}|(s_k^\tau)^\dag|\psi^{\sigma_n,\gamma}_{J,J^z_n}}
\braket{\psi^{\sigma,\gamma'}_{J',J^z_m-\tau}|\rho|\psi^{\sigma',\gamma'}_{J',J^z_n-\tau}} \Bigg\}
\label{eq:Lf}
\end{multline}
with the energy difference $\Delta^{\sigma,J,J^z}_{\sigma',J',{J^z}'} = \epsilon^\sigma_{J,J^z} - \epsilon^{\sigma'}_{J',{J^z}'}$.
Since the eigenenergy of the eigenstate is independent of the label $\gamma$, we combine these matrix elements into a $\gamma$-independent probability distribution $p^J_{J^z_m,\sigma_m;J^z_n,\sigma_n}$ using the degree of degeneracy, Eq.~\eqref{eq:Jdeg},
\begin{equation}
\begin{split}
p^J_{J^z_m,\sigma_m;J^z_n,\sigma_n} &= \sum_\gamma \braket{\psi^{\sigma_m,\gamma}_{J,J^z_m}|\rho|\psi^{\sigma_n,\gamma}_{J,J^z_n}} \\
&= g_N(J) \braket{\psi^{\sigma_m,\gamma}_{J,J^z_m}|\rho|\psi^{\sigma_n,\gamma}_{J,J^z_n}} .
\end{split}
\label{eq:p}
\end{equation}
Finally, using Eq.\ \eqref{eq:Lf}, we arrive at the rate equation,
\begin{multline}
\partial_t p^J_{J^z_m,\sigma_m;J^z_n,\sigma_n} =
-i \Delta^{\sigma_m,J,J^z_m}_{\sigma_n,J,J^z_n} p^J_{J^z_m,\sigma_m;J^z_n,\sigma_n} \\
- \Bigg\{ \sum_{\tau} \sum_{J',\sigma'} \left[
\Gamma^\tau_{J',J}(J^z_m+\tau,J^z_m+\tau;\sigma',\sigma',\sigma_m,\sigma_m) \right. \Bigg. \\
\Bigg. \left. + \Gamma^\tau_{J',J}(J^z_n+\tau,J^z_n+\tau;\sigma',\sigma',\sigma_n,\sigma_n) \right] \Bigg\}
p^J_{J^z_m,\sigma_m;J^z_n,\sigma_n} \\
+ \sum_\tau \sum_{J',\sigma,\sigma'} 2 \Gamma^\tau_{J,J'}(J^z_m,J^z_n;\sigma_m,\sigma_n,\sigma,\sigma') p^{J'}_{J^z_m-\tau,\sigma;J^z_n-\tau,\sigma'},
\label{eq:Lp}
\end{multline}
for $p^J_{J^z_m,\sigma_m;J^z_n,\sigma_n}$.
The prefactors for the three terms inducing transitions between the elements are combined into the total transition rate
\begin{multline}
\Gamma^\tau_{J,J'}(J^z_a,J^z_b;\sigma_a,\sigma_b,\sigma_c,\sigma_d) = 
\delta_{\epsilon^{\sigma_a}_{J,J^z_a},\epsilon^{\sigma_b}_{J,J^z_b}}
\delta_{\epsilon^{\sigma_c}_{J',J^z_a+\tau},\epsilon^{\sigma_d}_{J',J^z_b+\tau}}\\
\times\frac{1}{g_N(J')}
\sum_k W^\tau_k 
h^\tau_k(\Delta^{\sigma_a,J,J^z_a}_{\sigma_c,J',J^z_a-\tau}) \\
\sum_{\gamma,\gamma'}
\braket{\psi^{\sigma_a,\gamma}_{J,J^z_a}|s_k^\tau|\psi^{\sigma_c,\gamma'}_{J',J^z_a-\tau}}
\braket{\psi^{\sigma_d,\gamma'}_{J',J^z_b-\tau}|(s_k^\tau)^\dag|\psi^{\sigma_b,\gamma}_{J,J^z_b}} .
\label{eq:Grate}
\end{multline}

The occurring matrix elements $\braket{\psi^{\sigma,\gamma}_{J,J^z}|s_k^\tau|\psi^{\sigma',\gamma'}_{J',J^z-\tau}}$ for the spin operator $s_k^\tau$, Eq.~\eqref{eq:sktau}, are evaluated separately for the electron spin operator $S^\tau$ and the nuclear spin operator $I^\tau_k$.
Substitution of the explicit form of the eigenstates, Eq.~\eqref{eq:bm2s}, yields
\begin{multline}
\braket{\psi^{\sigma,\gamma}_{J,J^z}|S^\tau|\psi^{\sigma',\gamma'}_{J',J^z-\tau}} = 
\delta_{J,J'} \delta_{\gamma,\gamma'} \\
\times
\begin{cases}
c^{\sigma}_{J,J^z} d^{\sigma'}_{J',J^z-\tau} / \sqrt{2}, & \tau = -1, \\
(d^{\sigma}_{J,J^z} d^{\sigma'}_{J',J^z-\tau} - c^{\sigma}_{J,J^z} c^{\sigma'}_{J',J^z-\tau})/ 2, & \tau = 0, \\
d^{\sigma}_{J,J^z} c^{\sigma'}_{J',J^z-\tau} / \sqrt{2}, & \tau = +1,
\end{cases}
\label{eq:tmes}
\end{multline}
for the electron spin operator due to the orthonormality of the nuclear spin states.
For the nuclear spin operator we obtain the matrix elements
\begin{multline}
\braket{\psi^{\sigma,\gamma}_{J,J^z}|I_k^\tau|\psi^{\sigma',\gamma'}_{J',J^z-\tau}} = \\
c^{\sigma}_{J,J^z} c^{\sigma'}_{J',J^z-\tau} \braket{J,J^z,\gamma|I^\tau_k|J',J^z-\tau,\gamma'} \\
+ d^{\sigma}_{J,J^z} d^{\sigma'}_{J',J^z-\tau} \braket{J,J^z-1,\gamma|I^\tau_k|J',J^z-\tau-1,\gamma'}
\label{eq:tmei}
\end{multline}
as a result of the orthonormality of the electron spin states.

For the calculation of the remaining matrix elements of the type 
$\braket{J',J^z+\tau,\gamma'|I^\tau_k|J,J^z,\gamma}$, 
we make use of the assumption that the nuclear spins in the box model approximation are indistinguishable and, in compliance, omit any potential dependence of $W^\tau_k$ and $h^\tau_k$ on the individual nuclear spin $k\in \left\{1,\ldots,N\right\}$,
 i.e. we set $W^\tau_k = W^\tau_n$ and $h^\tau_k = h^\tau_n$ for all nuclear spins.
The electron spin contribution of these quantities, $W^\tau_0 = W^\tau_e$ and $h^\tau_0 = h^\tau_e$, however differs from that of the nuclear spins.
As a consequence of the assumption, the result of the evaluation for an individual nuclear spin $k$ can be adopted for the other nuclear spins as well, such that the nuclear contribution in the sum over $k$ in Eq.~\eqref{eq:Grate} solely produces a prefactor $N$.
The actual evaluation of the elements 
$\braket{J',J^z+\tau,\gamma'|I^\tau_k|J,J^z,\gamma}$ can be performed by virtue of the Clebsch-Gordan coefficients.
The results are presented in Appendix \ref{app:sfe}.

With the above considerations, the transition rate, Eq.~\eqref{eq:Grate}, can be transformed into
\begin{multline}
\Gamma^\tau_{J,J'}(J^z_a,J^z_b;\sigma_a,\sigma_b,\sigma_c,\sigma_d) = \\
\delta_{\epsilon^{\sigma_a}_{J,J^z_a},\epsilon^{\sigma_b}_{J,J^z_b}}
\delta_{\epsilon^{\sigma_c}_{J',J^z_a-\tau},\epsilon^{\sigma_d}_{J',J^z_b-\tau}}
\Big\{ W_e^0 h_e(\Delta^{\sigma_a,J,J^z_a}_{\sigma_c,J',J^z_a-\tau}) \Big. \\
\left. \times \braket{\psi^{\sigma_a,\gamma}_{J,J^z_a}|S^\tau|\psi^{\sigma_c,\gamma'}_{J',J^z_a-\tau}}
\braket{\psi^{\sigma_d,\gamma'}_{J',J^z_b-\tau}|(S^\tau)^\dag|\psi^{\sigma_b,\gamma}_{J,J^z_b}} \right. \\
\left. + N W_n^0 \sum_{j=J\pm1/2}\sum_{j'=J'\pm1/2} \frac{g_{N-1}(j')}{g_N(J')} 
h_n(\Delta^{\sigma_a,J,J^z_a}_{\sigma_c,J',J^z_a-\tau}) \right. \\
\Big. \times \braket{\psi^{\sigma_a,\gamma}_{J,J^z_a}|I_k^\tau|\psi^{\sigma_c,\gamma'}_{J',J^z_a-\tau}}
\braket{\psi^{\sigma_d,\gamma'}_{J',J^z_b-\tau}|(I_k^\tau)^\dag|\psi^{\sigma_b,\gamma}_{J,J^z_b}} \Big\}
\label{eq:Gratef}
\end{multline}
where the first term of the sum in the brace accounts for the electron spin flips and the second incorporates spin flips in the nuclear spin bath.
For the electron contribution, the flip rate $W_e^0$ is assumed to be independent on the sign of $\tau$, and the degree of degeneracy, $g_N(J')$, cancels out by the summation over $\gamma'$.
For the nuclear spin flips, we also introduced an isotropic rate $W_n^0$ identical for all nuclei.
The sums over $\gamma$, $\gamma'$ were treated as described in the Appendix~\ref{app:sfe} and yield sums over the quantum number $j$, $j'$ of the total nuclear spin's length in the reduced nuclear spin bath excluding the spin $k$ as well as the degree of degeneracy $g_{N-1}(j')$ as a prefactor.
The quantum numbers $j$, $j'$ are restricted to the values $j=J\pm1/2$, and $j'=J'\pm1/2$ respectively, which enter in the evaluation of the spin flip elements in the last line of Eq.~\eqref{eq:Gratef}, see Appendix \ref{app:sfe} for details.

The temperature-dependent function $h_{e,n}(\epsilon)$ entering the transition rates, Eq.~\eqref{eq:Gratef}, is chosen as
\begin{equation}
{h_{e,n}}(\epsilon) = \begin{cases} e^{-{\beta_{e,n}} \epsilon}, & \epsilon>0, \\
				1, & \epsilon \leq 0, \end{cases}
\label{eq:hk}
\end{equation}
in accordance with Ref.~\cite{fischer_prb2020}.
Any transition reducing the system's energy, $\epsilon<0$, or leaving the energy unchanged, $\epsilon=0$, occurs with maximum rate $W_{e,n}^0$, whereas transitions increasing the hyperfine energy are exponentially suppressed with increasing inverse spin temperature $\beta_{e,n}$.
Since the above choice fulfills the relation $h_{e,n}(\epsilon) / h_{e,n}(-\epsilon) = e^{-\beta_{e,n} \epsilon}$ it properly describes coupling with the thermal reservoirs with particular temperature.
Such a choice also ensures the correct Boltzmann weighted distribution of the steady-state density matrix in thermal equilibrium, $\beta_e = \beta_n$.

\section{Nuclear-spin polaron state}
\label{sec:polaronstate}

The Lindblad approach providing the steady-state density operator of the system for a broad temperature range, $T_e$ and $T_n$, forms the basis for the study of the crossover from the disordered high-temperature state to the correlated nuclear-spin polaron states in the low temperature regime.

\subsection{Electron-nuclear spin correlation functions. Anisotropy effects}
\label{sec:nuclear-polaron-correlators}

For the investigation of the nuclear-spin polaron formation, it is instructive to study the correlation of the charge carrier spin and the nuclear spins as shown in Ref.~\cite{fischer_prb2020} for the case of Ising coupling by comparing different criteria of nuclear spin polaron formation.
Indeed, the hyperfine energy of the system is minimized when the electron spin and the nuclear spins align in opposite directions and produce an anti-correlation of the electron and nuclear spins at a positive sign of the hyperfine coupling constants and the anisotropy parameter, i.e., at $A_0>0$ and $\lambda>0$.
In this case, the value of the electron-nuclear spin correlator will be negative.
If $A_0<0$, a positive correlation between the electron and nuclear spins is expected to form, i.e., the central spin and nuclear spin bath will be co-polarized.

However, the examination of the system at low temperatures reveals a profound dependence of the forming nuclear-spin polaron state on the anisotropy factor $\lambda$ of the hyperfine interaction, Eq.~\eqref{eq:bm}.
We illustrate the nature of the polaron state by the expectation value of the electron-nuclear spin correlation as a function of the inverse nuclear spin temperature $\bn$ at a fixed inverse electron spin temperature, $\be \wh = 0.5$.
For the density operator entering the calculation of the expectation value of an observable $O$, $\left<O\right> = \mr{Tr}\left[ O \rho \right]$, we insert the steady state solution $\rho_0$ of Eq.~\eqref{eq:Lp}.

\begin{figure}[t!]
\centering
\includegraphics[scale=1]{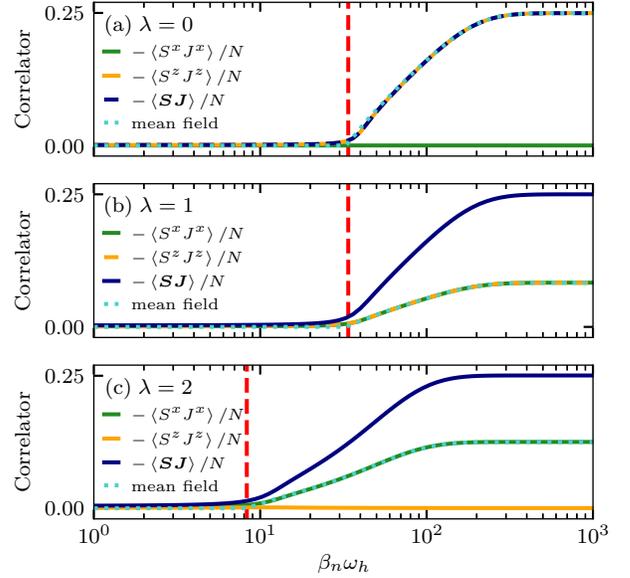}
\caption{Electron-nuclear spin correlation as a function of the inverse nuclear spin temperature $\bn$ for various anisotropy factors $\lambda$ of the hyperfine interaction. The inverse electron spin temperature is fixed at $\be \omega_h =0.5$. The dashed vertical red lines correspond to the transition temperatures according to the analytical Eq.~\eqref{eq:tt}. Mean-field results are added as turquoise dotted lines.}
\label{fig:sj}
\end{figure}

The data presented in Fig.~\ref{fig:sj} is obtained for a system with $N=1000$ nuclear spins in the box model approximation.
By varying the value of the hyperfine anisotropy $\lambda$, we selected the three physically particularly relevant cases: (a) the Ising case at $\lambda=0$ previously addressed in Ref.~\cite{fischer_prb2020}, (b) the isotropic case at $\lambda=1$, and (c) the case of the strong in-plane hyperfine coupling at $\lambda=2$.
For each case, we study the spacial components of the electron-nuclear spin correlator, $\left< S^x J^x \right>$ (green lines) and $\left< S^z J^z \right>$ (orange lines), separately as well as the total correlation $\left< \mathbf{S} \mathbf{J} \right>$ (blue lines).
The component $\left< S^y J^y \right>$ is not displayed since it is identical to $\left< S^x J^x \right>$ due to the axial rotation ($U(1)$) symmetry of the Hamiltonian, Eq.~\eqref{eq:bm}.
For the same reason, correlators of different spin components, $\langle S^\alpha J^\beta\rangle$ with $\alpha \ne \beta$, vanish.
The flip rates for the electron spin and the nuclear spins are set to $W_e^0 = 10^{-3} \omega_h$ and $W_n^0 = 10^{-6} \omega_h$ providing a three orders of magnitude faster flipping of the electron spin compared to the nuclear spins.
This choice of the rates and the number $N$ is kept throughout the whole work.

The overall behavior of the correlators as a function of the inverse nuclear spin temperature is similar for all three cases:
at high nuclear spin temperatures (small $\beta_n$) all correlators are negligible.
With a reduction of the nuclear spin temperature (increase in $\beta_n$) at least one correlator $\langle S^\alpha J^\alpha \rangle$ and the total spin correlator $\langle \bm S \bm J\rangle$ become significant.
They increase with increasing $\beta_n$ and at $T_n\to 0$ ($\beta_n \to \infty$) saturate.
However, as functions of the anisotropy parameter $\lambda$, the correlators of different electron-nuclear spin components demonstrate different behavior.

In the limit of $\lambda =0$ depicted in Fig.~\ref{fig:sj}(a), the hyperfine interaction consists solely of the Ising contribution along the $z$ axis.
The spin flip terms, i.e.\ the transversal hypefine contributions, are absent.
Therefore, the anti-correlation of the electron spin and the nuclear spins only builds up in $z$ direction, whereas the correlation functions of transversal components $\left< S^x J^x \right>=\langle S^y J^y\rangle$ remain zero.
At low temperatures (large $\be$ and $\bn$) the anti-correlation per nuclear spin reaches the maximum value $1/4$ determined by the product of the electron spin length and the spin length of an individual nuclear spin~\cite{fischer_prb2020}.
Since the coupling of the transversal components is absent in the Ising limit, the full correlator $\left< \mathbf{S} \mathbf{J} \right>$ is solely made up by the $z$ contribution.
Interestingly, a similar behavior is displayed by any system with an anisotropy factor in the range $0 \leq \lambda < 1$, for which the hyperfine interaction in $z$ direction is stronger than the $x$ and $y$ components.
Our calculations show that within the numerical accuracy for $\lambda \in [0,1)$ the results coincide with those shown in Fig.~\ref{fig:sj}(a).

In the isotropic case, $\lambda = 1$, see Fig.~\ref{fig:sj}(b), the nuclear-spin polaron state, that forms at large $\bn$, has different characteristics.
Due to the lack of spatial preference, the polaron state is isotropic:
The correlators $\left< S^x J^x \right>=\left< S^y J^y \right>$ and $\left< S^z J^z \right>$ build up equally with decreasing temperature.
As a result, the full correlator $\left< \mathbf{S} \mathbf{J} \right>$ is made up by equal contributions for the three spatial directions.
At low nuclear spin temperatures ($\beta_n\to \infty$) it reaches $\left< \mathbf{S} \mathbf{J} \right>/N =-1/4$, whereas each spatial component contributes with the value $-1/12$.

An anisotropy factor $\left| \lambda \right| > 1$ is relevant, e.g., for light holes in QDs, where $\lambda = -2$ \cite{testelin_prb2009,hackmann_prl2015}.
Since the sign of $\lambda$ does not change the overall behavior of the system but affects the sign of the transversal electron-nuclear spin correlator only, i.e., it determines whether the electron spin and the nuclear spins align parallel or anti-parallel within the $(xy)$ plane, we restrict ourselves to positive values of $\lambda$.
The results for $\lambda=2$ are depicted in Fig.~\ref{fig:sj}(c).
Here the transversal contributions of the hyperfine interaction dominate over the $z$ contribution.
Thus, an anti-correlation of the electron and nuclear spins builds within the $(xy)$ plane while no (anti-)correlation in $z$ direction arises.
Consequently, the total anti-correlation, $-\left< \mathbf{S} \mathbf{J} \right>,$  is split between the $x$ and $y$ component which have a maximum value of $1/6$ per nuclear spin.
Note that for $\left| \lambda \right| > 1$ the crossover regime, where the nuclear-spin polaron state starts to form (indicated by the dashed vertical lines in Fig.~\ref{fig:sj}) is shifted to higher temperatures.
This effect is discussed in more detail in Sec.~\ref{sec:temperature-crit} below.

\subsection{Mean-field approach to the anisotropic system}
\label{sec:mf}

For a deeper understanding of the nuclear-spin polaron state that forms in a spin system with anisotropic hyperfine coupling, we refer to a mean-field approach which previously was developed by Merkulov for the isotropic system \cite{merkulov_pss1998}.
In the mean-field approximation we assume the electron spin to experience the average effective field generated by the nuclear spins,  i.\ e., the average Overhauser field $\left< \mathbf{B_N} \right>$, caused by the nuclear spin polarization.
In their turn, the nuclear spins are subject to the average effective field of the electron spin, the average Knight field $\left< \mathbf{B_K} \right>$.
These effective fields result in the polarization of the respective spin systems in the form
\begin{subequations}
\begin{align}
\left< \mathbf{S} \right> &= - \frac{\left< \mathbf{B_N} \right>}{2 \left| \left< \mathbf{B_N} \right> \right|} \tanh \left( \frac{\beta_e \left| \left< \mathbf{B_N} \right> \right|}{2} \right), \label{eq:mfs} \\
\left< \mathbf{J} \right> &= - \frac{N \left< \mathbf{B_K} \right>}{2 \left| \left< \mathbf{B_K} \right> \right|} \tanh \left( \frac{\beta_n \left| \left< \mathbf{B_K} \right> \right|}{2} \right) , \label{eq:mfj}
\end{align}
\end{subequations}
where the definitions of the Overhauser field and the Knight field include the anisotropy parameter $\lambda$ of the hyperfine interaction
\begin{subequations}
\begin{align}
\left< \mathbf{B_N} \right> &= A_0 \left( \lambda \left< J^x \right>, \lambda \left< J^y \right>, \left< J^z \right> \right)^T , \label{eq:bn} \\
\left< \mathbf{B_K} \right> &= A_0 \left( \lambda \left< S^x \right>, \lambda \left< S^y \right>, \left< S^z \right> \right)^T , \label{eq:bk}
\end{align}
\end{subequations}
and the fields are measured in the energy units.

To obtain the self-consistency equation for the total nuclear spin $\left< \mathbf{J} \right>$, Eq.~\eqref{eq:mfs} is inserted into Eq.~\eqref{eq:mfj} taking into account the definitions of $\left< \mathbf{B_N} \right>$ and $\left< \mathbf{B_K} \right>$,
\begin{multline}
\left< \mathbf{J} \right> = \frac{N}{2 L_1} 
\tanh \left[ \frac{\beta_n A_0}{4} \frac{L_2}{L_1}
\tanh \left( \frac{\beta_e A_0}{2} L_1 \right) \right] \\
\times \left( \lambda \left< J^x \right>, \lambda \left< J^y \right>, \left< J^z \right>\right)^T \label{eq:mfsc}
\end{multline}
where we introduced $L_1 = \sqrt{\lambda^2 ( \left< J^x \right>^2 + \left< J^y \right>^2) + \left< J^z \right>^2}$ and $L_2 = \sqrt{\lambda^4 ( \left< J^x \right>^2 + \left< J^y \right>^2) + \left< J^z \right>^2}$ for brevity.

In order to obtain the critical temperature of the polaron formation let us denote
the angle between the vector $\langle \mathbf J\rangle$ and the $z$ axis by $\theta\in [0,\pi]$.
Since the system is isotropic in the $(xy)$ plane the polar angle of $\langle \mathbf J\rangle$ is unimportant.
As a first step we solve Eq.~\eqref{eq:mfsc} for the absolute value $\left| \left< \mathbf{J} \right> \right|$ and obtain that the polaron can be formed in the mean-field approach provided that the following condition
\begin{equation}
\frac{NA_0^2 \beta_e \beta_n}{16} \sqrt{\lambda^4 \sin^2 \theta + \cos^2 \theta } > 1
\label{eq:mfcond}
\end{equation}
is fulfilled.
Thus, the parameter $\lambda$ induces a modification of the critical temperatures especially for angles $\theta$ close to $\pi/2$.

As a next step we determine the orientation of the spins in the polaron by solving the self-consistency equation for the angle $\theta$.
It can be derived from Eq.~\eqref{eq:mfsc} using the relation $\tan^2 \theta = ( \left< J^x \right>^2 + \left< J^y \right>^2 ) / \left< J^z \right>^2$ and taking into account that the left and right hand sides of Eq.~\eqref{eq:mfsc} should be parallel:
\begin{equation}
\tan^2 \theta = \lambda^4 \tan^2 \theta .
\label{eq:mfsctheta}
\end{equation}
Equation \eqref{eq:mfsctheta} reveals the potential orientations of the polaron state with respect to $\lambda$.
We find that in the isotropic case, $\lambda=1$, the relation holds for arbitrary $\theta$.
Otherwise Eq.~\eqref{eq:mfsctheta} is only consistent with three solutions for the angle $\theta$:
$\theta = 0$, $\theta = \pi$, or $\theta = \pi/2$.
A stability analysis, see Appendix \ref{app:mfstability}, demonstrates that for $\lambda < 1$ the states with $\theta = 0$ and $\theta = \pi$ are stable and $\theta = \pi/2$ is an unstable solution, whereas for $\lambda > 1$ the categorization is switched, i.e., $\theta = \pi/2$ is stable and $\theta = 0$, $\theta = \pi$ are not.
Thus, the mean-field calculations predict that the nuclear-spin polaron forms along the $z$ axis for $\lambda<1$ (easy-axis situation) and within the $(xy)$ plane for $\lambda>1$ (easy-plane situation).
As a result, the polaron formation condition within the mean-field approach can be summarized as:
\begin{equation}
\label{mean:field:fin}
\frac{NA_0^2 \beta_e \beta_n}{16} >
\begin{cases}
1, \quad |\lambda|\leqslant 1,\\
\lambda^{-2}, \quad |\lambda|>1.
\end{cases}
\end{equation}
Naturally, the symmetry breaks in such a way that polarizations build up to maximize the absolute value of the hyperfine coupling.
This analysis is consistent with the results obtained above, in Sec.~\ref{sec:nuclear-polaron-correlators}.

The mean-field solutions for the electron-nuclear spin correlation $\left< S^\alpha J^\alpha \right>/N$ (with $\alpha\in \left\{x,y,z \right\}$) for those spatial components $\alpha$, in which the anti-correlation builds in the low-temperature regime, are added in Fig.~\ref{fig:sj} (dotted turquoise lines) alongside the data obtained by our approach as a comparison.
We find that within the presented temperature range, the two approaches nearly coincide.
The mean-field solution, however, exhibits a sharper transition to the polaron state at the critical temperature consistent with a phase transition even in non-equilibrium, while a smooth crossover is observed in the finite system, see Ref.~\cite{fischer_prb2020} for more details.

\subsection{Crossover temperature for the polaron formation}
\label{sec:temperature-crit}

The mean-field approach, Eq.~\eqref{mean:field:fin}, predicts the formation of a nuclear-spin polaron state below the critical temperatures given by
\begin{equation}
\beta_{e,c} \beta_{n,c} = \frac{16}{N \tilde{A}_0^2} .
\label{eq:mfct}
\end{equation}
The equation combines the criteria for the polaron state along the $z$ direction ($\theta=0$/$\theta=\pi$) and for the polaron oriented within the $(xy)$ plane ($\theta=\pi/2$) by introducing a rescaled hyperfine coupling constant
\begin{equation}
\tilde{A}_0= \begin{cases} A_0, & \lambda \leq 1, \\
			\lambda A_0, & \lambda > 1. \end{cases}
\label{eq:tildea}
\end{equation}

In Ref.~\cite{fischer_prb2020} we derived a more complex temperature criterion for the polaron-state formation based on the rate-equation formalism taking into account the finite number of nuclear spins.
We substitute the coupling constant $\tilde{A}_0$ into Eq.\ (31) of Ref.\ \cite{fischer_prb2020} and obtain the temperature criterion
\begin{equation}
\beta_{n,t} = \frac{4}{\tilde{A}_0} \artanh \left( \frac{4}{(N+2) \beta_{e,t} \tilde{A}_0} \right)
\label{eq:tt}
\end{equation}
for the onset of polaron formation generalized to an arbitrary anisotropy.
This defines a line in the $(\beta_{n},\beta_{e})$ plane.

\begin{figure}[t!]
\centering
\includegraphics[scale=1]{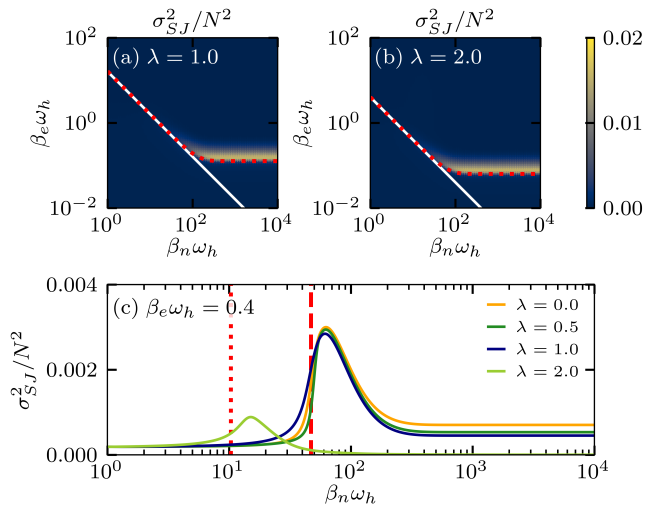}
\caption{Fluctuations $\sigma_{SJ}^2$ of the electron-nuclear spin correlator, Eq.~\eqref{fluct:noise}, as a function of the effective inverse nuclear spin temperature $\beta_n$ and the effective inverse electron spin temperature $\beta_e$ for different values of the hyperfine anisotropy parameter, (a) $\lambda=1$ and (b) $\lambda=2$, and (c) at a fixed electron spin temperature, $\beta_e \omega_h = 0.4$.}
\label{fig:sigmasj}
\end{figure}

As a common indicator for the crossover to the nuclear-spin polaron state for all values of the hyperfine parameter $\lambda$, we focus on the total electron-nuclear spin correlation since we found that $\left< \mathbf{S}\mathbf{J}\right>$ is maximized consistently in the polaron state,  cf. Fig.~\ref{fig:sj}.
The crossover temperature line extracted from the master equation approach is then indicated by the rise of the fluctuations of $\left< \mathbf{S}\mathbf{J}\right>$,
\begin{equation}
\label{fluct:noise}
\sigma^2_{SJ} = \left< (\mathbf{S}\mathbf{J})^2 \right> - \left< \mathbf{S}\mathbf{J} \right>^2,
\end{equation}
which we plotted as a color contour plot in the $(\beta_{n},\beta_{e})$ plane for $\lambda=1$ in Fig.~\ref{fig:sigmasj}(a) and for $\lambda=2$ in Fig.~\ref{fig:sigmasj}(b).
The temperature line defined in Eq.~\eqref{eq:tt} (depicted as a red dotted line) matches the line formed by the maximum of $\sigma^2_{SJ}$.
For comparison, the mean-field critical temperature, Eq.~\eqref{eq:mfct}, is added as well (white line).

For the physical interpretation of the fluctuations $\sigma^2_{SJ}$ we refer to case of equal temperatures $\beta_e = \beta_n$:
At low temperatures, the spins are aligned either within the $(xy)$ plane or in $z$ direction (depending on $\lambda$), and the hyperfine energy is proportional to the spin correlator $\left< \mathbf{S}\mathbf{J}\right>$.
Therefore the fluctuations of the correlator in thermal equilibrium are proportional to the heat capacity of the system which is expected to display a discontinuity at the critical temperature in the Landau theory of phase transitions \cite{landau_lifshitz_book1}.
Since we consider a finite system with $N=1000$ nuclear spins here, the system does not exhibit a genuine phase transition but a crossover behavior that becomes sharper with increasing $N$.
The peak in the fluctuations $\sigma^2_{SJ}$ as a function of $\beta_n$, see Fig.~\ref{fig:sigmasj}(c), is relatively sharp, and its rising edge is positioned at the crossover temperature according to Eq.~\eqref{eq:tt} (red dashed vertical line for $\lambda\leq1$, red dotted vertical line for $\lambda=2$).
It is noteworthy that for $\lambda=2$ the peak of the fluctuations $\sigma^2_{SJ}$ at a fixed electron temperature, $\beta_e \omega_h = 0.4$, is less pronounced than for $\lambda\leq1$ due to the shift of the polaron regime to lower temperatures when $\lambda>1$.

\subsection{Nuclear distribution functions}
\label{sec:nucldistfunc}

\begin{figure*}[t!]
\centering
\includegraphics[scale=1]{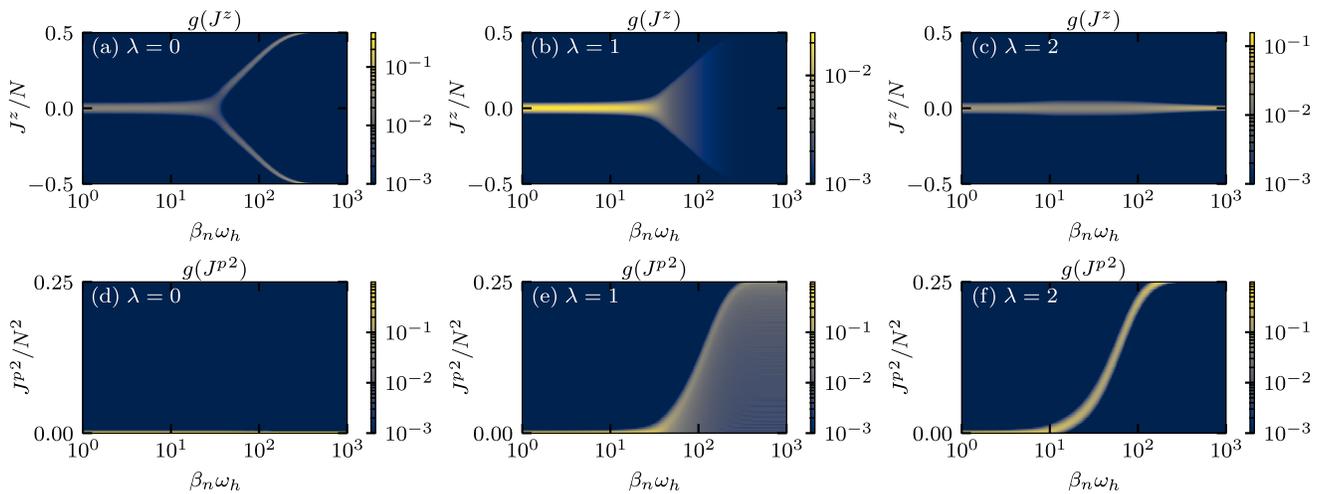}
\caption{Distribution function of the nuclear spin quantum numbers $J^z$ (upper panels) and ${J^p}^2$ (lower panels) for three typical values of the anisotropy factor $\lambda$ of the hyperfine interaction with $N=1000$ nuclear spins. The inverse nuclear spin temperature $\bn$ is displayed on the horizontal axis; the inverse electron spin temperature is fixed at $\be \wh =0.5$.}
\label{fig:gjxz}
\end{figure*}

Aiming at a comprehensive investigation of the polaron formation beyond the mean-field approach, we focus on the distribution functions of the nuclear spin quantum numbers which provide an ideal tool to study the reorientation of the nuclear spins related to the formation of a nuclear-spin polaron state in the cooled system.
To this end, we consider again the steady-state density operator of Eq.~\eqref{eq:Lp} at a given electron and nuclear spin temperature and define the distribution function,
\begin{eqnarray}
g(J^z) &=& \sum_{J,\sigma} (c^\sigma_{J,J^z})^2 p^J_{J^z,\sigma;J^z,\sigma}
\nonumber \\
&&  + \sum_{J,\sigma}(d^\sigma_{J,J^z+1})^2 p^J_{J^z+1,\sigma;J^z+1,\sigma}
\end{eqnarray}
by transforming from the energy eigenbasis into the spin $z$ basis and summing all contributions with a fixed nuclear spin quantum number $J^z$.
In addition, we introduce the quantum number of the perpendicular component of the total nuclear spin,
\begin{equation}
{J^p}^2 = J(J+1) - {J^z}^2 ,
\end{equation}
that is deduced from the quantum numbers $J$ and $J^z$ and is restricted to ${J^p}^2 \in \left\{0,\dots,N/2(N/2+1)\right\}$.
The related distribution function,
\begin{multline}
g({J^p}^2) = \sum_{J,J^z,\sigma} \left[ (c^\sigma_{J,J^z})^2 p^J_{J^z,\sigma;J^z,\sigma} \right. \\
\left. + (d^\sigma_{J,J^z+1})^2 p^J_{J^z+1,\sigma;J^z+1,\sigma} \right] \delta_{{J^p}^2, (J(J+1)-{J^z}^2)} ,
\end{multline}
is obtained by summation of all contributions to a given value of ${J^p}^2$ analogously to $g(J^z)$.
To display the distribution function $g({J^p}^2)$, the data is processed into a histogram with appropriate bin size (typically $100$ bins within the range ${J^p}^2 \in \left[0,N/2(N/2+1)\right]$).

The distribution functions, $g(J^z)$ and $g({J^p}^2)$, as a function of the effective inverse nuclear spin temperature $\beta_n$ for fixed $\beta_e \omega_h = 0.5$ are displayed in Fig.~\ref{fig:gjxz}.
In the high-temperature limit (small $\beta_n$), the nuclear spins are randomly aligned and
$J^z$ follows an approximately Gaussian distribution centered around zero independent on the hyperfine parameter $\lambda$.
In the high-temperature limit the nuclear spin system is isotropic.
Hence, the distribution of ${J^p}^2$ at high temperatures is proportional to $\exp (-2{J^p}^2/N)$.
However when decreasing the temperature (increasing $\beta_n$) the distributions are altered below a certain point:
The behavior of the system is now determined by the hyperfine interaction and its anisotropy.

In the Ising limit, where the nuclear-spin polaron state is oriented along the positive/negative $z$ direction, we find the two possible orientations reflected by two branches forming for $g(J^z)$, depicted in Fig.~\ref{fig:gjxz}(a).
These results fully match the data in Ref.~\cite{fischer_prb2020} obtained by the kinetic rate equations taking into account the diagonal elements of the density operator.
Naturally, the ${J^p}^2$ component remains distributed closely around zero at $\lambda=0$, see Fig.~\ref{fig:gjxz}(d).

\begin{figure}[t]
\centering
\includegraphics[scale=1]{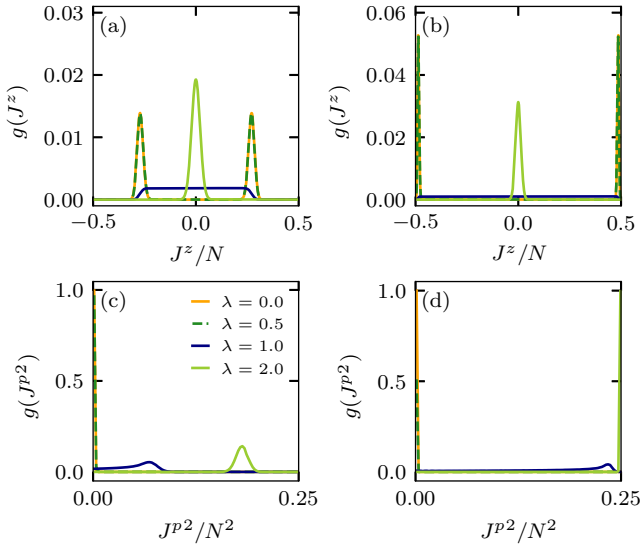}
\caption{Distribution function of the nuclear spin quantum numbers $J^z$ (upper panels) and ${J^p}^2$ (lower panels) for various anisotropy factors $\lambda$ of the hyperfine interaction, see legend in panel (c). The inverse nuclear spin temperature is set to $\bn \wh = 80$ [panels (a) and (c)] or $\bn \wh = 270$ [panels (b) and (d)]; the inverse electron spin temperature is fixed at $\be \omega_h =0.5$.}
\label{fig:gjxzbn}
\end{figure}

For a better illustration the vertical cut through the panels of Fig.~\ref{fig:gjxz}, is displayed in Fig.~\ref{fig:gjxzbn} for two different values of $\beta_n$.
Here the two peaks in $g(J^z)$ (yellow lines for $\lambda=0$) move further apart with increasing $\beta_n$ from $\bn \wh = 80$ in the left hand panels to $\bn \wh = 270$ in the right hand panels.
As a comparison we added the data for $\lambda=0.5$ in Fig.~\ref{fig:gjxzbn} as well.
In this case we find similar behavior as for the Ising limit though the peak of $g({J^p}^2)$ at $\bn \wh = 270$ is a bit broader indicating that for $0<\lambda<1$ certain correlations appear also in the in-plane nuclear spin components.

For the system with isotropic hyperfine interaction, the distributions $g(J^x)$, $g(J^y)$ and $g(J^z)$ coincide, see Fig.~\ref{fig:gjxz}(b) and Figs.~\ref{fig:gjxzbn}(a) and \ref{fig:gjxzbn}(b) (blue lines) for $g(J^z)$ as an example.
The narrow Gaussian distribution of $J^z$ of the high-temperature regime transforms into a wide and almost uniform distribution at low temperatures.
The range of the uniform distributions broadens with decreasing the temperature until the full range $J^z \in \left[ -N/2, N/2\right]$ is covered.
The distributions for $\lambda=1$ at fixed nuclear spin temperatures, see Fig.~\ref{fig:gjxzbn} (blue lines), are nearly flat.
The uniform distribution of the quantum number $J^z$ complies with the uniform distribution of the polaron orientation on the Bloch sphere.
Accordingly the distribution of ${J^p}^2$ is roughly given by $g({J^p}^2) = 1/(J(J+1)-{J^p}^2)^{1/2}$ at low temperatures, see Fig.~\ref{fig:gjxz}(e) and \ref{fig:gjxzbn}(d).

In an anisotropic system with $\lambda>1$, the polaron forms within the $(xy)$ plane.
The nuclear distribution functions reflect this fact by narrowing the distribution $g(J^z)$ around $J^z=0$ when lowering the temperature starting from the initial Gaussian distribution.
This is depicted in Fig.~\ref{fig:gjxz}(c) as well as Fig.~\ref{fig:gjxzbn}(a) and Fig.~\ref{fig:gjxzbn}(b) for $\lambda=2$.
At the same time the weight in the distribution of ${J^p}^2$ moves from ${J^p}^2=0$ to the maximum value ${J^p}^2=N/2(N/2+1)$ resulting from the maximum quantum number $J=N/2$ and the minimum value $J^z=0$.

\subsection{Quantum phase transition}
\label{sec:quantum-phase-transition}

The dependence of the nuclear-spin polaron state on the hyperfine anisotropy parameter $\lambda$ also tracks the transition of the ground state of the Hamiltonian, Eq.~\eqref{eq:bm}, at a critical coupling $\lambda_c=1$.

The lowest eigenenergy of the eigenenergies stated in Eqs.~\eqref{eq:bm1e} and \eqref{eq:bm2e} 
is always given by Eq.~\eqref{eq:bm2e} for the case $\sigma=-$ and a maximum value of $J$, i.e., $J=N/2$ (we recall that we consider $N$ to be even), independent on the parameter $\lambda$ and takes the form
\begin{align}
\epsilon^-_{J,J^z} &= - \frac{A_0}{4} - \frac{A_0}{2} \Bigg\{ \frac14 + \lambda^2  J(J+1) \Bigg. \notag \\
& \qquad \Bigg. + \left( 1 - \lambda^2 \right) J^z \left(J^z - 1 \right) \Bigg\}^{1/2} .
\end{align}
We note that replacing $J^z$ by $-(J^z-1)$ results in the same eigenenergy.

In the above formulation it becomes clear that for $\lambda^2 < 1$ the term $J^z(J^z-1)$ has to maximize, and therefore the ground states results from $J^z=-N/2+1$ or $J^z=N/2$.
At $\lambda =1$, the value of $J^z$ does not influence the eigenenergy, and the ground state is $N$-fold degenerate in $J^z$.
By contrast, for $\lambda^2 > 1$, the ground state requires a minimum of the term $J^z(J^z-1)$, which corresponds to $J^z = 0$ or $J^z=1$.
Therefore, the system undergoes a quantum phase transition at $\lambda_c\equiv1$ with a change of the ground state degeneracy from a twofold degenerate ground state for $|\lambda|<1$ or $|\lambda|>1$ to a degeneracy of $N$ for $|\lambda|=\lambda_c$.
For odd $N$ the degeneracy of the ground state is $1$ for $|\lambda|>\lambda_c$.

The difference between the two ground states for $\lambda^2<1$ and the two ground states for $\lambda^2>1$ lies in the fact that for $\lambda^2<1$ there is no transition between the two ground states via single spin-flip processes which disconnects these ground states for $\lambda^2<1$.
Hence at zero temperature, thermal spin-flip processes between the two ground states are inhibited.
For $\lambda^2 > 1$ however the two ground states with $J^z = 0$ and $J^z=1$ are directly connected by a single spin-flip process.
The coupling to the environment provides non-zero transition matrix elements as stated in Eqs.~\eqref{eq:tmes} and \eqref{eq:tmei} such that even at zero temperature fluctuations between the two ground states will take place.
In the mean-field approach, presented in Sec.~\ref{sec:bm}, the difference in the nature of the ground state translates to two disconnected polaron states for $\lambda^2<1$ whereas the nuclear spin polaron forms isotropically within the $(xy)$ plane for $\lambda^2>1$.

\begin{figure}[t!]
\centering
\includegraphics[scale=1]{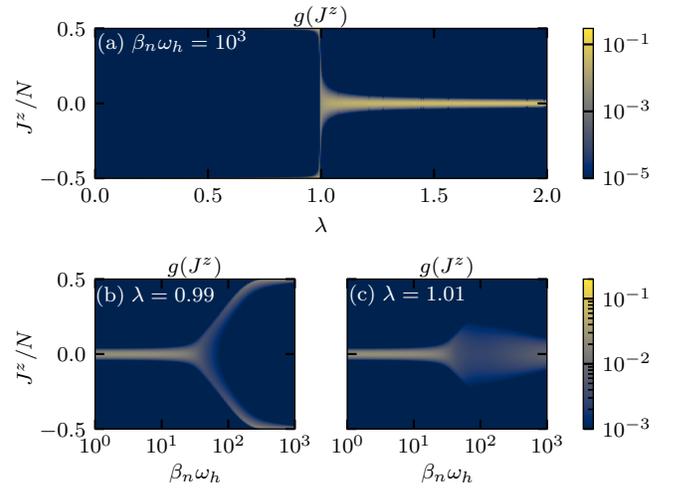}
\caption{Distribution function of the nuclear spin quantum number $J^z$. (a) Dependence on the anisotropy factor $\lambda$ of the hyperfine interaction for fixed inverse spin temperatures, $\bn \wh = 1000$, $\be \omega_h =0.5$. (b) and (c) Dependence on the inverse nuclear spin temperature with $\lambda$ adjusted slightly below (b) or above one (c); $\be \omega_h =0.5$.}
\label{fig:gjxzlambda}
\end{figure}

The quantum phase transition at $\lambda_c=1$ translates to a rapid change of the nuclear distribution function $g(J^z)$ at low temperatures, see Fig.~\ref{fig:gjxzlambda}(a).
For $\lambda<1$, the distribution function has two very sharp peaks at $J^z/N = \pm 0.5$ (which therefore are hard to detect in Fig.~\ref{fig:gjxzlambda}(a)), whereas for $\lambda>1$ the distribution displays a single maximum around $J^z=0$.
In the isotropic limit, $\lambda=1$, $g(J^z)$ covers the full range of potential values of $J^z$ uniformly.
Around the point of isotropy we find a blurred behavior as a result of the finite non-zero temperatures.

The nuclear spin distribution functions for systems close to the quantum critical point reveal that even a slight anisotropy leads to a well distinguished signature of both phases at low temperatures.
We picked $\lambda=0.99<\lambda_c$ and $\lambda=1.01>\lambda_c$ as an example
and plotted the temperature evolution of the distribution function in Fig.~\ref{fig:gjxzlambda}(b) and Fig.~\ref{fig:gjxzlambda}(c) respectively.

For $\lambda=0.99$, the two polaron branches corresponding to the opposite spin alignments in $z$ direction appear similar to Fig.~\ref{fig:gjxz}(a).
In comparison to the results in the Ising limit $\lambda=0$, the branches are just slightly broadened at intermediate inverse nuclear spin temperature $\beta_n$.

The data for $\lambda=1.01$ exhibits similar deviations from the case of stronger anisotropy $\lambda=2$ in Fig.~\ref{fig:gjxz}(c), whereas the overall sharpening of the nuclear distribution function $g(J^z)$ around $J^z=0$ is the same.
For $\lambda=1.01$, however, $g(J^z)$ first resembles the isotropic case in the regime of intermediate temperatures close to the crossover temperature resembling the distribution depicted in Fig.~\ref{fig:gjxz}(b).
Only with further decreasing of the nuclear spin temperatures the distribution focuses around $J^z=0$.

Note that the anisotropy factor for the hyperfine interaction of electron spins in semiconductor nanostructures equates to the quantum critical point $\lambda_c=1$.
Derivation from an isotropic system are characteristic for localized hole spins and significantly effect the polaron formation.

\section{Temporal spin fluctuations}
\label{sec:fluct}

Nuclear-spin polaron formation strongly affects the temporal dynamics of the electron and nuclear spin degrees of freedom.
The direct access to it is provided by the time-dependent spin correlation functions.
In this section we study electron and nuclear spin fluctuations in time domain and highlight the role of the nuclear-spin polaron effects.

\subsection{Electron spin fluctuations}
\label{sec:szsz}

The temporal fluctuations $\left<S^z(0)S^z(t)\right>$ of the electron spin are accessible by optical measurements of the electron spin noise~\cite{aleksandrov81,Oestreich:rev,smirnov:SNS:rev}.
In terms of the Lindblad-master equation formalism, Eq.~\eqref{eq:La}, the electron spin fluctuations are calculated by the quantum mechanical trace with the steady-state density operator $\rho_0$,
\begin{equation}
\begin{split}
C_S^z(t) = \left< S^z(0) S^z(t) \right> &= \tr \left[ \rho_0 S^z S^z(t) \right] \\
&= \tr \left[ S^z e^{\mathcal{L}t} (S^z \rho_0) \right],
\end{split}
\label{eq:szszt}
\end{equation}
where $\mathcal{L}$ is the Liouvillian operator determining the time evolution of the open quantum system and the superoperator $\exp(\mathcal{L}t)$ is applied to $S^z \rho_0$ \cite{Lax1963}.

\begin{figure}[t]
\centering
\includegraphics[scale=1]{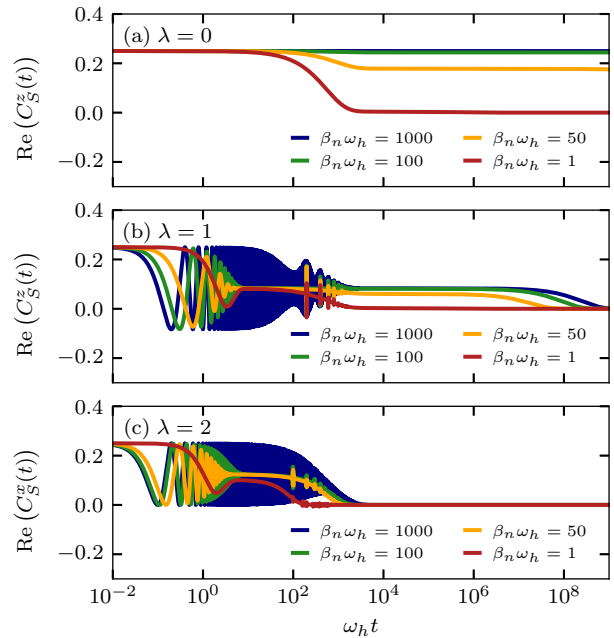}
\caption{Temporal fluctuations of the electron spin components for different values of the hyperfine anisotropy parameter $\lambda$ in (a), (b) and (c). Results for various effective inverse nuclear spin temperatures $\beta_n$ are presented respectively whereas $\beta_e \omega_h = 0.5$ is kept constant.}
\label{fig:szszt}
\end{figure}

Figure \ref{fig:szszt} presents the electron spin autocorrelation as a function of time for three distinct values of the hyperfine anisotropy parameter $\lambda$.
The initial value of the electron spin correlator yields $C_S^z(0)=1/4$ regardless of the temperature since both electron spin components are equiprobable.
The electron spin at low temperatures displays long living correlations, related to the spin polaron formation, whose lifetime depends on the choice of $\lambda$, whereas in the high-temperature regime the autocorrelation function completely decays on a timescale given by the inverse thermal electron spin flip rate $\tau_s =1/W_e^0$ ($ = 10^{3}/ \omega_h$ for our choice of parameters) demonstrating also nontrivial dynamics at shorter timescales.

In the Ising limit, $\lambda=0$, the correlator decays to zero on a timescale proportional to $\tau_s$ at high temperatures, see Fig.~\ref{fig:szszt}(a) (red line).
In this situation the hyperfine interaction does not effect the electron spin-$z$ component and its decay is fully controlled by the reservoir induced spin-flip processes.
However, when the effective nuclear spin temperature is reduced to the crossover temperature where the polaron formation sets in, the correlator $C_S^z(t)$ does not decay completely anymore within the presented time range up to $t\omega_h=10^9$ but to a plateau with a finite non-zero value (orange line).
The degree of correlation at this plateau increases with the lowering of temperatures.
At very low temperatures, e.g., $\beta_n \omega_h = 1000$ (blue line) deep in the polaronic phase, no decay is visible anymore, and the full correlation of the electron spin persists for the full time interval presented in the figure.
With lowering the temperatures, the reservoir induced spin-flip processes become more and more suppressed, cf.\ Eq.~\eqref{eq:hk}, which shifts the decay of $C_S^z(t)$ to longer time scales.
However, we expect a decay of $C_S^z(t)$ to zero on a prolonged time scale for non-zero temperatures as a result of the exponentially suppressed but non-zero flip rates.

In the isotropic system, the spin-flip terms of the hyperfine Hamiltonian come into play and yield a two-stage behavior.
In the high-temperature limit, the electron spin initially dephases in the nuclear spin bath with the rate $\omega_h$ which produces the characteristic curve $C_S^z(t)$ that reaches a plateau of the value $1/12$, see Fig.~\ref{fig:szszt}(b) (red line), analytically derived in Refs.~\cite{KT,merkulov_prb2002} in the limit of frozen nuclear spins for a closed system.
However, the correlator decays further on a time scale determined by the inverse rate $1/W_e^0$ due to the coupling to the thermal reservoirs~\cite{gi2012noise}.

Note that the equidistant spikes in the correlators for $\lambda=1$ and $\lambda=2$
at time scales of $ \omega_h t\approx 10^2-10^3$ in Fig.~\ref{fig:szszt}(b,c)
are an artifact of the box model approximation and the finite number $N$ of nuclear spins.
For equal hyperfine coupling constants $A_k = A_0$ of all nuclear spins, the Overhauser field is quantized, i.e., the spacial components in Eq.~\eqref{eq:bn} can only assume values that are an integer multiple of $\lambda A_0$.
Thus the precession frequencies of the electron spin are all commensurate and yield a rephasing at times $T_n = 2 \pi n/ (\lambda A_0)$ with an integer $n\in \left\{ -N/2,\ldots,N/2\right\}$~\cite{glazov_book}.

A reduction of the effective nuclear spin temperature yields an oscillatory component to $C_S^z(t)$ in the isotropic system in the absence of an external magnetic field, since the electron spin starts to precess around the emerging nuclear spin polarization which is isotropically distributed and therefore contains components perpendicular to the $z$ axis.
Lowering the temperature, the nuclear spins become more and more oriented and generate a stronger Overhauser field such that the electron precession frequency increases.

Additionally, the stronger nuclear alignment reduces the fluctuations of the nuclear spin which prevents the dephasing of the electron spin and results in an elongated envelope of the oscillating $C_S^z(t)$.
At times $t \gtrsim 1/W_e^0$, the electron spin flip processes resulting from the coupling to the thermal reservoir come into play and provide further dephasing such that the oscillatory component eventually vanishes even at low temperatures and $C_S^z(t)$ reaches the plateau of $1/12$ \cite{KT,merkulov_prb2002} which stems from the spatial electron spin component parallel to the Overhauser field and is protected from thermal spin flips due to a large energy barrier.
The plateau persists for several orders of magnitude in time and then decays further on a timescale determined by the effective nuclear spin temperature and the electron and nuclear spin flip rates in the system.

This decay can be attributed to the rotation of the nuclear-spin polaron state.
Since the system is fully isotropic, a polaron state, for which exemplarily the electron spin formerly was aligned in $z$ direction, may rotate such that the electron spin points in any other direction on the Bloch sphere.
Thus the temporal correlation of the electron spin $z$ component will get lost.
The rate of this loss of correlation may be understood by means of a diffusion process on the diagonal of $S^z \rho_0$ entering Eq.~\eqref{eq:szszt}, see Appendix~\ref{app:rotation} for details.
As a consequence the total rate for the rotation of the nuclear spin polaron state is approximately made up by
\begin{equation}
W_r = W_e^0/N^2 + W_n^0/N .
\label{eq:wr}
\end{equation}

For rates $W_e^0 = 10^{-3} \omega_h$, $W_n^0 = 10^{-6} \omega_h$ and $N=1000$, the rate for rotation of the polaron state correspondingly is $W_r=2\times10^{-9}\omega_h$ which matches the low-temperature result in Fig.~\ref{fig:szszt}(b) (blue line).
The rotation of the nuclear polaron state for $\lambda=1$ maintains a finite rate $W_r$ even for zero temperatures, hence the correlations exhibit a fundamental difference from the case $\lambda<1$ that originates from the different nature of the ground states.

The auto-correlation function of the electron spin $x$ component, $C_S^x(t) = \left< S^x(0) S^x(t) \right>$, for the system with a hyperfine anisotropy $\lambda=2$ resembles the results for the isotropic system and is plotted in Fig.~\ref{fig:szszt}(c).
Due to the amplification of the hyperfine interaction in $x$ and $y$ direction, the dip in the high-temperature limit predicted by Ref.~\cite{KT,merkulov_prb2002} for the isotropic case is shifted to earlier times.
The correlation function for the spin $z$ component, $C_S^z(t)$, for $\lambda=2$ alongside the spin $x$ component, $C_S^x(t)$, for $\lambda=0$ is provided in App.~\ref{app:cperp} for completeness.

Naturally, the electron spin precession is faster for $\lambda=2$ than for $\lambda=1$ at the same temperatures due to the enhanced Overhauser field perpendicular to the $z$ axis.
The correlator $C_S^x(t)$ for $\lambda=2$ decays on the timescale dictated by the rate $W_e^0$ of thermal electron spin flips even in the low-temperature limit such that the electron spin correlations are limited to a lifetime of $10^3\omega_h$ for our choice of parameters:
The twofold degenerate (non-degenerate) ground state does not protect the electron spin correlator from the dephasing induced by thermal electron spin flip processes.
In other words, it is a consequence of the in-plane isotropy of the system.

\subsection{Fluctuations of the nuclear spins}
\label{sec:jzjz}

\begin{figure}[t!]
\centering
\includegraphics[scale=1]{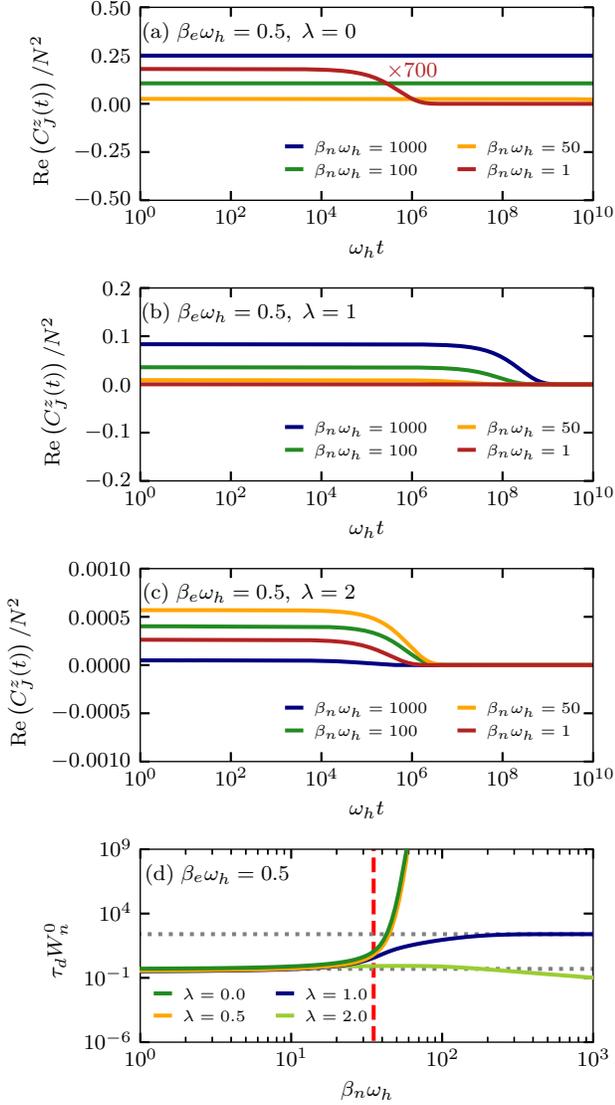}
\caption{Temporal fluctuations of the nuclear spin $z$ component for (a) the Ising limit, (b) the isotropic system, and (c) the anisotropic system with $\lambda=2$. The inverse electron spin temperature $\beta_e$ is fixed; the inverse nuclear spin temperature $\beta_n$ is encoded by different colors.
(d) Decay time $\tau_d$ of the nuclear spin fluctuations for different values of the hyperfine anisotropy parameter $\lambda$. The transition temperature, Eq.~\eqref{eq:tt}, is added as a red dashed vertical line.}
\label{fig:jzjzt}
\end{figure}

The long living correlations of the electron spin at low temperatures are related to the similar dynamics of the nuclear spin bath.
In contrast to the electron spin, however, the nuclear correlator does not display any fast modulations but is constant for a long time until a temperature-dependent decay process may take place.
Figure~\ref{fig:jzjzt} displays the nuclear correlator,
\begin{equation}
\begin{split}
C^z_J(t)= \left< J^z(0) J^z(t) \right> &= \tr \left[ \rho_0 J^z J^z(t) \right] \\
&= \tr \left[J^z e^{\mathcal{L}t} (J^z \rho_0 ) \right] ,
\end{split}
\label{eq:jzjzt}
\end{equation}
at different inverse nuclear spin temperatures, Fig.~\ref{fig:jzjzt}(a) for the Ising limit, Fig.~\ref{fig:jzjzt}(b) in the isotropic system, and Fig.~\ref{fig:jzjzt}(c) for the anisotropic system with $\lambda=2$.

Since the spin-fluctuation induced by the thermal reservoirs will ultimately cause a decay to zero for $t\to\infty$ at non-zero temperatures, we define the decay time $\tau_d$ as the point in time where the correlator has reduced by the fraction $e$ with respect to its initial value, i.\ e.\ $C^z_J(\tau_d)= C^{z}_J(0)/e$.
We plot $\tau_d$ as a function of the effective inverse nuclear spin temperature $\beta_n$ whereas the electron spin temperature, $\beta_e \omega_h = 0.5$, remains constant.
The data for various values of the hyperfine anisotropy parameter $\lambda$ is presented in Fig.~\ref{fig:jzjzt}(d).

At high temperatures the decay is inherently dictated by the thermal nuclear spin flip rate $W_n^0$ (in our calculations, $W_n^0=10^{-6}\omega_h$) independent on the hyperfine anisotropy $\lambda$.
The related decay time $1/2W_n^0$ is indicated in Fig.~\ref{fig:jzjzt}(d) by the lower horizontal dotted grey line.

Moving to the temperature regime of the nuclear-spin polaron formation, the characteristic decay time of the correlator $C^{z}_J(t)$ increases for $\lambda<1$ similar to the electron spin correlator depicted in Fig.~\ref{fig:szszt}(a).
At low temperatures, $C^{z}_J(t)$ does not reach half of the starting value within our largest simulation time of $t=10^{15}/\omega_h$, see Fig.~\ref{fig:jzjzt}(a).
In the context of the quantum phase transition, we pointed out that the two-fold degenerate ground state maximizes $J=N/2$ as well as $J^z$ so that the spin flips induced by the thermal reservoir become exponentially suppressed leading to an exponential increase in $\tau_d$.
Therefore, the decay time $\tau_d$ grows exponentially starting at the transition temperature, Eq.~\eqref{eq:tt}, see red dashed vertical line in Fig.~\ref{fig:jzjzt}(d).

For $\lambda \geq 1$, this exponential increase of $\tau_d$ is absent as a result of the rotational symmetry in the nuclear-spin polaron state.
In the isotropic system, the orientation of the nuclear-spin polaron state rotates with the rate $W_r$ ($W_r=2\times10^{-9}\omega_h$ for our choice of parameters) previously deduced in the considerations of the electron spin correlation, see Eq.~\eqref{eq:wr}.
Accordingly, $\tau_d$ in the temperature range of polaron formation rises to approximately $1/2W_r = 0.25 \times 10^{9} \omega_h$ (upper horizontal dotted grey line in Fig.~\ref{fig:jzjzt}(d)).

For $\lambda>1$ the decay time $\tau_d$ reduces when the nuclear spin temperature is lowered.
For an explanation we refer to the dynamic rotation of the polaron state in the isotropic case.
Here, the dynamics of the non-zero matrix-elements of the composite operator $O=J^z \rho_0$ in the groundstate at $T_e=T_n=0$ follow Eq.~\eqref{eq:chioffdiagonal} (off-diagonal elements) and Eq.~\eqref{eq:chidiagonal} (diagonal elements) respectively.
The differential equations yield a decoupled decay of the off-diagonal elements with approximate rate $W_e^0+NW_n^0$, while transitions between the diagonal elements occur with the same rate $W_e^0+NW_n^0$.
The ground state for $\lambda>1$ is solely two-fold degenerate in contrast to the $N$-fold degeneracy in the isotropic case, cf.~Sec.~\ref{sec:quantum-phase-transition}.
Thus, for $\lambda>1$, a single spin flip between the two ground states (generating a transition between the two non zero diagonal elements of $(J^z \rho_0)$) already leads to a complete loss of correlation whereas in the isotropic case the correlation is gradually lost by successive spin flips.
As a result, the decay of the correlator $C^{z}_J(t)$ for $\lambda>1$ remains bound to the decay rate $\tau_d \approx (W_e^0+NW_n^0)^{-1}$ when reducing the temperature, while in the isotropic system the decay is prolonged in the polaronic state.

\section{Conclusion}
\label{sec:conclusion}

We generalized the kinetic approach for the nuclear-polaron formation to an arbitrary anisotropic CSM.
This allows us to investigate all experimentally relevant regimes of singly charged QDs and localized electronic carriers.
We proposed a symmetry conserving Lindblad approach that is applicable to arbitrary hyperfine coupling anisotropy factors $\lambda$ and calculated the steady-state solution for two distinct reservoir temperatures $T_e$ and $T_n$.
Our approach overcomes the limitation of Ref.~\cite{fischer_prb2020} to $\lambda=0$ but includes the previously investigated limit as well.

We have studied the electron-nuclear spin correlator, the nuclear spin distribution function and the temporal autocorrelators of the spins.
The spin correlation functions as well as the nuclear distribution function reveal the nuclear polaronic state formation when reducing the nuclear spin temperature.
The crossover temperature into the nuclear polaron state coincides with enhanced fluctuations of the spin-correlation function and also agrees with a mean-field theory prediction for the anisotropic CSM.

Importantly, we demonstrate a quantum phase transition at the anisotropy parameter $\lambda=1$ which separates distinct polaronic states.
For $\lambda <1$ the result in the polaronic phase is identical to the Ising limit:
spin fluctuations are suppressed by a very large activation barrier.
At $\lambda=1$ the polaron state is fully rotationally invariant, while for $\lambda>1$ we find a rotational invariant phase around the $z$ axis.

Our approach makes it possible to study not only the steady state of the electron-nuclear spin system, but also the dynamics of the polaron formation and temporal fluctuations of spins.

\begin{acknowledgments}

We acknowledge financial support by the Deutsche Forschungsgemeinschaft and the Russian Foundation of Basic Research through the transregio TRR 160 within the Projects No.\ A4, and No.\ A7.
M.M.G. was partially supported by RFBR-DFG project No. 19-52-12038.
The authors gratefully acknowledge the computing time granted by the John von Neumann Institute for Computing (NIC) under Project HDO09 and provided on the supercomputer JUWELS at the J\"ulich Supercomputing Centre.
\end{acknowledgments}

\appendix

\section{Details on the level degeneracy}
\label{app:degen}

The idea of separating the degeneracy factors $g(\epsilon_{n,m})$ in the transition rates $\Gamma_{m,n}^{k,\alpha}$, Eq.~\eqref{eq:gamma-rate}, becomes clear when inserting the Lindblad operator, Eq.~\eqref{eq:Lb}, into the Eq.~\eqref{eq:La}
\begin{align}
\dot{\rho} &=
-i \left[ H, \rho \right] 
- \sum_{k,\alpha} \sum_{m,n} \sum_{a,b} \sum_{a',b'}
\frac{W_k^\alpha h_k^\alpha(\Delta_{mn})}{g(\epsilon_m) g(\epsilon_n)} \notag\\
&\quad \times \delta_{\epsilon_m,\epsilon_a} \delta_{\epsilon_n,\epsilon_b}
\delta_{\epsilon_m,\epsilon_{a'}} \delta_{\epsilon_n,\epsilon_{b'}} \notag\\
& \quad \times \left\{ \delta_{a',a}
\left( \left( s_k^\alpha \right)_{a',b'} \right)^\dag
\left(s_k^\alpha\right)_{a,b} 
(X_{b'b} \rho +\rho X_{b'b}) \right. \notag\\
&\qquad \left. -2 \left( s_k^\alpha \right)_{a,b}
\left( \left(s_k^\alpha \right)_{a',b'} \right)^\dag
X_{ab}\rho X_{b' a'} \right\} .
\label{eq:Lc}
\end{align}
We abbreviated  the matrix elements of the spin operators by $ \left( s_k^\alpha \right)_{a,b}=\braket{\psi_a|s_k^\alpha|\psi_b}$ and exploited the orthonormality of eigenstates, $\braket{\psi_{a'}|\psi_a} = \delta_{a',a}$.
Due to the relation 
\begin{eqnarray}
\sum_{m,n} \delta_{\epsilon_m,\epsilon_a} \delta_{\epsilon_n,\epsilon_b} &= &g(\epsilon_a) g(\epsilon_b)
\end{eqnarray}
the levels of degeneracy cancel out of the equation,
\begin{multline}
\dot{\rho} = -i \left[ H, \rho \right]
- \sum_{k,\alpha} \sum_{a,b} \sum_{a',b'} W_k^\alpha h_k^\alpha(\Delta_{ab}) \\
\times \delta_{\epsilon_a,\epsilon_{a'}} \delta_{\epsilon_b,\epsilon_{b'}}
\times \left\{ \ldots \right\} ,
\label{eq:Ld}
\end{multline}
where the term within the brace remains unchanged as in Eq.~\eqref{eq:Lc} and therefore is abbreviated by "$\ldots$". 
This clarifies why we introduced the degeneracy factors in the definition of the Lindblad rates $\Gamma_{m,n}^{k,\alpha} $ above.

\section{Spin flip matrix elements}
\label{app:sfe}

As a first step for evaluating $\braket{J',J^z+\tau,\gamma'|I^\tau_k|J,J^z,\gamma}$ we disentangle the quantum number $\gamma$ that accounts for the degeneracy of the $J$ quantum number.
Since we are interested in flipping an individual nuclear spin, the state $\ket{J,J^z,\gamma}$ is cast into the format $\ket{J,J^z,j,\gamma_j,1/2}$.
Here $j$ labels the quantum number of total nuclear spin length excluding the spin $k$ (whose length is indicated by the $1/2$ in the notation) and can take on the values $j= J\pm1/2$.
The quantity $\gamma_j$ is the equivalent of $\gamma$ in the reduced nuclear spin bath without spin $k$, i.e. $\gamma_j$ accounts for the degeneracy of $j$ in a spin bath of size $N-1$.
Consequently, the sum over $\gamma$ ($\gamma'$) in the transition rate, Eq.~\eqref{eq:Grate}, is split into a sum over the quantum numbers $j$ ($j'$) and $\gamma_j$ ($\gamma_j'$) where the latter simply produces a factor of degeneracy $g_{N-1}(j)$ ($g_{N-1}(j')$) according to the definition, Eq.~\eqref{eq:Jdeg}.
For brevity, the indices $\gamma_j$, $\gamma_j'$ are omitted in the following notation.

In the former summations, the contributions $j=J\pm1/2$, $j'=J'\pm1/2$ have to be evaluated individually.
To this end, a state is disassembled into states of format $\ket{j,j^z;1/2,I_k^z}$ according to
\begin{multline}
\ket{J,J^z,j=J\pm1/2,1/2} = \\
\mp \sqrt{\frac12 \left( 1 \mp \frac{J^z}{j +1/2} \right)} \ket{j,J^z-1/2;1/2,1/2} \\
+ \sqrt{\frac12 \left( 1 \pm \frac{J^z}{j + 1/2} \right)} \ket{j,J^z+1/2;1/2,-1/2} .
\end{multline}
Here, we can eventually apply the nuclear spin operator $I_k^\tau$ which yields
\begin{subequations}
\begin{gather}
I^{+1}_k \ket{J,J^z,j=J\pm 1/2,1/2} = \notag\\
\frac{1}{2} \sqrt{\left( 1 \pm \frac{J^z}{j + 1/2} \right)}
\ket{j,J^z+1/2;1/2,1/2} \\
I^{0}_k \ket{J,J^z,j=J\pm 1/2,1/2} = \notag \\
\mp \sqrt{\frac18 \left( 1 \mp \frac{J^z}{j +1/2} \right)} \ket{j,J^z-1/2;1/2,1/2} \notag \\
- \sqrt{\frac18 \left( 1 \pm \frac{J^z}{j + 1/2} \right)} \ket{j,J^z+1;1/2,-1/2} \\
I^{-1}_k \ket{J,J^z,j=J\pm 1/2,1/2} = \notag \\
\mp \frac12 \sqrt{ \left( 1 \mp \frac{J^z}{j +1/2} \right)}
\ket{j,J^z-1/2;1/2,-1/2} .
\end{gather}
\end{subequations}
For the elements $\braket{J',J^z+\tau,j',1/2|I_k^\tau|J,J^z,j,1/2}$ one obtains consequently
\begin{subequations}
\begin{gather}
\braket{J',J^z+1,j'=J'\pm1/2,1/2|I_k^{+1}|J,J^z,j=J\pm1/2,1/2} \notag\\
=\mp \delta_{j,j'} \frac{1}{2} \sqrt{\frac12 \left( 1 \pm \frac{J^z+1}{j + 1/2} \right) \left( 1 \mp \frac{J^z}{j +1/2} \right)} \\
\braket{J',J^z,j'=J'\pm1/2,1/2|I_k^{0}|J,J^z,j=J\pm1/2,1/2} \notag\\
= \delta_{j,j'} \frac14 \left\{ \sqrt{\left( 1 \mp \frac{J^z}{j +1/2} \right)^2} - \sqrt{\left( 1 \pm \frac{J^z}{j + 1/2} \right)^2} \right\} \\
\braket{J',J^z-1,j'=J'\pm1/2,1/2|I_k^{-1}|J,J^z,j=J\pm1/2,1/2} \notag\\
= \mp \frac12 \sqrt{\frac12 \left( 1 \mp \frac{J^z}{j +1/2} \right) \left( 1 \pm \frac{J^z-1}{j + 1/2} \right)} .
\end{gather}
\end{subequations}

\section{Stability analysis for mean field solutions}
\label{app:mfstability}

When the parameter $\lambda \neq 1$, Eq.~\eqref{eq:mfsctheta} holds true for either $\theta = 0$, $\theta = \pi$, or $\theta = \pi/2$.
For the former case we reformulate the equation,
\begin{equation*}
\theta = \arctan \left( \pm \lambda^2 \tan \theta \right)
\end{equation*}
and perform a Taylor expansion for small angles
\begin{equation*}
\theta \approx \pm \lambda^2 \theta .
\end{equation*}
Insertion of a small perturbation $\Delta$ to the fix point $\theta =0$ yields that the point is stable when $\lambda^2 < 1$ and unstable when $\lambda^2 > 1$.
Similar results are obtained for the point $\theta =\pi$.
For the latter case, $\theta = \pi/2$, we consider an alternative version of the self-consistency equation for $\theta$.
To this end, we use $\cot^2 \theta = \left< J^z \right>^2 / ( \left< J^x \right>^2 + \left< J^y \right>^2 )$ and obtain
\begin{align*}
\cot^2 \theta &= \lambda^{-2} \cot^2 \theta \\
\theta &= \arccot \left( \pm \lambda{-2} \cot \theta \right) .
\end{align*}
The Taylor expansion for a small perturbation $\Delta$ around the angle $\pi/2$ yields
\begin{equation*}
\pi/2 + \Delta \approx \pi/2 \pm \lambda^{-2} \Delta .
\end{equation*}
Thus, the solution $\theta= \pi/2$ is stable for $\lambda^{-2} <1$, i.e. $\lambda^2 > 1$, and unstable for $\lambda^2 < 1$.

\section{Rotation of the nuclear spin polaron state}
\label{app:rotation}

The rate of the rotation of the nuclear spin polaron state in the isotropic system, and thereby the time scale of the decay of the correlation functions $C_S^z(t)$ and $C_J^z(t)$, can be derived from the rate equations for the elements of $O\rho_0$ that enters the definitions Eqs.~\eqref{eq:szszt} and \eqref{eq:jzjzt}.
The operator $O$ either corresponds to the operator of the electron spin, $S^z$, or the total nuclear spin, $J^z$.
For the matrix elements of $O\rho_0$ in the energy eigenbasis we introduce
\begin{align}
\begin{split}
\chi^J_{J^z_m,\sigma_m;J^z_n,\sigma_n} &= g_N(J) \braket{\psi^{\sigma_m,\gamma}_{J,J^z_m}|O\rho_0|\psi^{\sigma_n,\gamma}_{J,J^z_n}} \\
&= p^J_{J^z_n,\sigma_n;J^z_n,\sigma_n} \braket{\psi^{\sigma_m,\gamma}_{J,J^z_m}|O|\psi^{\sigma_n,\gamma}_{J,J^z_n}}
\end{split}
\end{align}
analogously to $p^J_{J^z_m,\sigma_m;J^z_n,\sigma_n}$ in Eq.~\eqref{eq:p}.
In the second line, we made use of the fact that the steady-state density operator $\rho_0$ is diagonal in the energy eigenbasis.

The time-dependent matrix elements $\chi^J_{J^z_m,\sigma_m;J^z_n,\sigma_n}(t)$ obey the same differential equation as $p^J_{J^z_m,\sigma_m;J^z_n,\sigma_n}$, see Eq.~\eqref{eq:Lp}.
Since we are interested in the dynamics of the polaron state at low temperatures, we refer to the limit $T_e=T_n=0$ for simplicity in the following.
At zero temperatures solely the diagonal elements $p^J_{J^z_n,\sigma_n;J^z_n,\sigma_n}$ in the subspace with $J=N/2$ and $\sigma_n=-$ are occupied as they constitute the ground state, cf.\ Sec.~\ref{sec:quantum-phase-transition}.
Furthermore, the operator $O$ fulfills the relations $\left[ \mathbf{J}^2, O\right] = \left[ J^z+S^z, O\right]=0$ and as a result does not generate transitions between energy eigenstates with distinct index $J$ or $J^z$ (actually corresponding to the total spin $z$ component).
Consequently, only the elements of type $\chi^{N/2}_{J^z,\sigma;J^z,-}$ have non-zero value.

We use the rate equation, Eq.~\eqref{eq:Lp}, separately for off-diagonal elements, $\chi^{N/2}_{J^z,+;J^z,-}$, and diagonal elements, $\chi^{N/2}_{J^z,-;J^z,-}$, respectively, to obtain their temporal evolution.
For the off-diagonal elements, transitions to other elements drop out since the rate $\Gamma^\tau_{J,J'}(J^z,J^z;+,-,\sigma,\sigma')$ in the last line of Eq.~\eqref{eq:Lp} vanishes.
The remaining terms in Eq.~\eqref{eq:Lp},
\begin{multline}
\dot{\chi}^{N/2}_{J^z,+;J^z,-} = - \chi^{N/2}_{J^z,+;J^z,-} \Big\{ i \Delta^{+,J,J^z}_{-,J,J^z} \Big. \\
\Big. + \sum_\tau \sum_{J',\sigma'} \left[ \Gamma^\tau_{J',J}(J^z+\tau,J^z+\tau;\sigma',\sigma',+,+) \right. \Big. \\
\Big. \left. + \Gamma^\tau_{J',J}(J^z+\tau,J^z+\tau;\sigma',\sigma',-,-) \right] \Big\} ,
\label{eq:chioffdiagonal}
\end{multline}
generate oscillations with the frequency $\Delta^{+,J,J^z}_{-,J,J^z} = A_0(J+1/2)$ that decay with a rate given by the sum over the bracket.
The resulting decay rate is roughly proportional to $W_e^0+NW_n^0$.
This approximation results from Eq.~\eqref{eq:Gratef} minding $T_e=T_n=0$ in the function $h_k(\Delta)$, evaluating $g_N(J=N/2)=g_{N-1}(j=N/2-1/2)=1$ and approximately setting the matrix elements $\braket{\psi^{\sigma_a,\gamma}_{J,J^z_a}|s^\tau_k|\psi^{\sigma_c,\gamma'}_{J',J^z_a-\tau}}$ to a constant.

For the diagonal elements $\chi^{N/2}_{J^z,-;J^z,-}$, the differential equation, Eq.~\eqref{eq:Lp}, simplifies to
\begin{multline}
\dot{\chi}^{N/2}_{J^z,-;J^z,-} = \\
2 \sum_\tau \left\{ -\Gamma^\tau_{N/2,N/2}(J^z+\tau,J^z+\tau;-,-,-,-) {\chi}^{N/2}_{J^z,-;J^z,-} \right. \\
\left. + \Gamma^\tau_{N/2,N/2}(J^z,J^z;-,-,-,-) \chi^{N/2}_{J^z-\tau,-;J^z-\tau,-} \right\} 
\label{eq:chidiagonal}
\end{multline}
where the sum over $J'$, $\sigma$, $\sigma'$ reduces to a single contribution when solely the ground states, $J'=N/2$ and $\sigma=\sigma'=-$, are occupied.
In the above equation the two terms for $\tau=0$ cancel out such that only the contributions $\tau = \pm 1$ remain.
The rates according to Eq.~\eqref{eq:Gratef} read
\begin{multline}
\Gamma^\tau_{J,J}(J^z,J^z;-,-,-,-) = \\
W_e^0 \braket{\psi^{-,\gamma}_{J,J^z}|S^\tau|\psi^{-,\gamma'}_{J,J^z-\tau}} \braket{\psi^{-,\gamma'}_{J,J^z-\tau}|(S^\tau)^\dag|\psi^{-,\gamma}_{J,J^z}} \\
+ NW_n^0 \braket{\psi^{-,\gamma}_{J,J^z}|I_k^\tau|\psi^{-,\gamma'}_{J,J^z-\tau}} \braket{\psi^{-,\gamma'}_{J,J^z-\tau}|(I_k^\tau)^\dag|\psi^{-,\gamma}_{J,J^z}}
\end{multline}
with $J=N/2$ and $J^z$ shifted to $J^z+\tau$ for the first term in the rate equation, Eq.~\eqref{eq:chidiagonal}.
Due to $T_e=T_n=0$, the function $h_k(\Delta)$ in the definition, Eq.~\eqref{eq:Gratef}, simplifies to a factor of one as does the degree of degeneracy $g_N(J=N/2)=g_{N-1}(j=N/2-1/2)=1$.
For simplicity the matrix elements of the spin flip operators are approximated by $1/\sqrt{8}$ respectively minding Eq.~\eqref{eq:sktau}.
As a consequence the rate equation reduces to
\begin{multline}
\dot{\chi}^{N/2}_{J^z,-;J^z,-} = \frac14 \left( W_e^0 + N W_n^0 \right)\\
\times \left( -2 \chi^{N/2}_{J^z,-;J^z,-} + \chi^{N/2}_{J^z+1,-;J^z+1,-} + \chi^{N/2}_{J^z-1,-;J^z-1,-} \right) .
\end{multline}

Employing the continuum limit for $J^z$ valid for $N\rightarrow\infty$ and replacing ${\chi}^{N/2}_{J^z,-;J^z,-}$ by the continuous function $\chi(J^z,t)$, the rate equation can be rewritten as
\begin{equation}
\partial_t \chi(J^z,t) = D \partial_{J^z}^2 \chi(J^z,t)
\end{equation}
with $D= \left( W_e^0 + N W_n^0 \right)/4$.

This corresponds to a diffusion equation which has the fundamental solution
\begin{equation}
\chi(J^z,t) = \frac{1}{\sqrt{4 \pi D t}} \exp \left( - {J^z}^2/ 4Dt \right)
\end{equation}
in 1D.
To obtain the characteristic rate of the rotation of the nuclear spin polaron state, we consider the standard deviation $\sigma_\chi$ of the above Gaussian and request $\sigma_\chi^2 = (N/2)^2$ for the diffusion process of the diagonal elements $\chi(J^z,t)$.
We obtain the relation $2Dt =  (N/2)^2$ where we insert the inverse rotation rate, $t = 1/2W_r$.
(The factor $2$ here stems from the definition of the prefactors in the Lindblad equation, Eq.~\eqref{eq:La}.)
Finally the rate of polaron rotation, Eq.~\eqref{eq:wr}, results.

\section{Fluctuations of the transversal electron spin component}
\label{app:cperp}

\begin{figure}[t]
\centering
\includegraphics[scale=1]{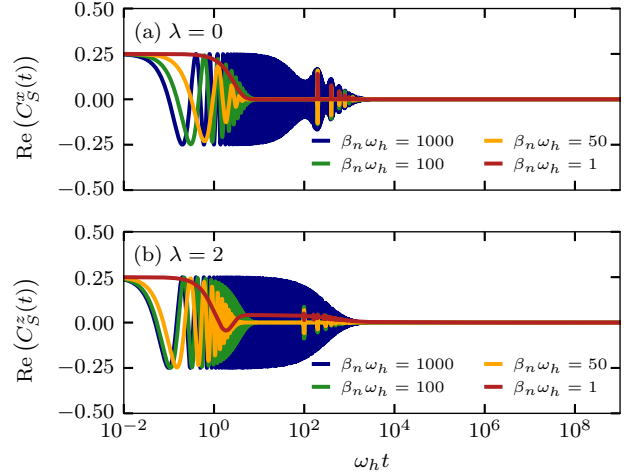}
\caption{Temporal fluctuations of the electron spin components perpendicular to the directions favored by hyperfine interaction for a hyperfine anisotropy parameter (a) $\lambda=0$ and (b) $\lambda=2$. Results for various effective inverse nuclear spin temperatures $\beta_n$ are presented respectively whereas $\beta_e \omega_h = 0.5$ is kept constant.}
\label{fig:szszt_perp}
\end{figure}

The temporal fluctuations of the electron spin along the spatial directions which are not favored by the anisotropic hyperfine interaction are presented in Fig.~\ref{fig:szszt_perp} for completeness.

In the Ising limit, $\lambda=0$, the hyperfine interaction acts along the $z$ axis only.
The autocorrelation function of the transversal electron spin component $C^x_S(t)=C^y_S(t)$ is presented in Fig.~\ref{fig:szszt_perp}(a) for various inverse nuclear spin temperatures.
We find an oscillatory component that builds up with decreasing the effective nuclear spin temperature and can be attributed to the polaron formation along the $z$ axis.
The envelope at high temperatures (red/orange curve) results from the electron spin dephasing in the disordered nuclear spin bath with a rate $\omega_h$.
At low temperatures, when the nuclear spins are oriented along the $z$ axis, the electron spin dephases on a prolonged time scale determined by the thermal electron spin flips with rate $W_e^0$.

In the anisotropic case, $\lambda=2$, the hyperfine interaction within the $(xy)$ plane is stronger than along the $z$ direction.
Here, the auto correlation $C^z_S(t)$ in the high-temperature limit, see Fig.~\ref{fig:szszt_perp}(b) (red curve), is slightly modified as compared to the predictions in the isotropic case \cite{KT,merkulov_prb2002} as a result of the anisotropy.
Additionally the thermal electron spin flips introduce a decay of $C^z_S(t)$ with the rate $W_e^0$.
At low temperatures the orientation of the nuclear spins within the $(xy)$ plane leads to oscillations in $C^z_S(t)$.
Again the dephasing rate changes from $\omega_h$ at high temperatures to $W_e^0$ in the low temperature regime.


%


\end{document}